\documentclass[11pt,a4paper]{article}
\usepackage{imakeidx}
\makeindex[columns=2, title=Alphabetical Index, intoc]
\usepackage{jheppub} 
\usepackage{lineno}
\usepackage{amsmath}
\usepackage{dsfont} 
\usepackage{nicefrac}
\usepackage[dvipsnames]{xcolor}
\usepackage{braket}
\usepackage{enumitem}
\usepackage{mathtools}
\usepackage{cleveref}
\usepackage{bbold}
\usepackage{physics}
\usepackage{graphicx}

\usepackage{multirow}
\usepackage{caption}
\usepackage{makecell}
\usepackage{subcaption}

\usepackage{tikz-feynman}
\usepackage{mathdots}
\usepackage{textgreek}

\usepackage{rotating}
\usepackage{cancel}
\usepackage{slashed}

\usepackage{longtable} 

\newcommand{\be}{\begin{equation}}
\newcommand{\ee}{\end{equation}}
\makeatletter
\newcommand\ps@indexpagestyle{
 \renewcommand\@oddfoot{\hfill-- \thepage\ --\hfill}
 \renewcommand\@oddhead{}
}
\makeatother
\indexsetup{firstpagestyle=indexpagestyle}

\def\KITITP{Institute for Theoretical Physics, KIT, Wolfgang-Gaede-Straße 1, 76131, Karlsruhe, Germany}
\def\KITA{Institute for Theoretical Particle Physics, KIT, Wolfgang-Gaede-Straße 1, 76131, Karlsruhe, Germany}

\preprint{
\begin{flushright} 
KA-TP-07-2026\\
TTP26-010 \\
P3H-26-021
\end{flushright}
}

\title{\boldmath  
A perturbative framework  to probe infrared sensitivity in non-Abelian gauge theories}

\author[a]{Duarte Fontes,}
\author[b]{Dennis Horstmann,}
\author[b]{Kirill Melnikov,}
\author[b]{Davide Maria Tagliabue}

\emailAdd{duarte.fontes@kit.edu}
\emailAdd{dennis.horstmann@kit.edu}
\emailAdd{kirill.melnikov@kit.edu}
\emailAdd{davide.tagliabue@kit.edu}

\affiliation[a]{\KITITP}
\affiliation[b]{\KITA}

\abstract{Understanding the infrared sensitivity of perturbative predictions in QCD is important for assessing the magnitude of possible non-perturbative power corrections to processes with large momentum transfer. In renormalon models, this sensitivity can be related to computable dependences of perturbative quantities on a small gluon mass. However, this procedure cannot be applied to collider processes with gluons at the Born level.
To address this problem, we promote the gluon mass to a parameter of a consistent non-Abelian quantum field theory where the gauge symmetry is spontaneously broken through the Higgs mechanism. Working in the limit in which the gluon mass $\mg$ is the smallest dimensionful parameter, we compute through two loops the $\order{\mg}$ contributions to the relation between the pole and $\overline{\rm MS}$ masses of a heavy quark and to the relation between corresponding field counterterms. We expect that the proposed framework will provide a useful laboratory for probing linear infrared sensitivity of collider observables in QCD.
}


\usepackage{bbding}
\usepackage{mathtools}
\usepackage{booktabs}
\usepackage{lscape}
\usepackage{url}
\usepackage{bbold}
\usepackage{bm}
\usepackage{csquotes}

\usepackage{xspace}
\usepackage{makecell}
\usepackage{xifthen}
\usepackage{placeins}

\usetikzlibrary{decorations.markings}
\tikzset{
  ghostArrow/.style={
    ghost,
    postaction={decorate},
    decoration={
      markings,
      mark=at position 0.6 with {\arrow{Stealth[length=7pt,width=5pt]}}
    }
  }
}

\usepackage{pgfplots}
\usepgfplotslibrary{colorbrewer}
\pgfplotsset{compat=newest}
\pgfplotsset{colormap/violet}

\usepackage{subcaption}
\usepackage{feynmp-auto}
\usepackage{color}
\definecolor{mycolor}{RGB}{165,111,29}
\usepackage[bottom]{footmisc}

\usepackage{dowith}

\newcommand{\UpperCase}[1]{
  \expandafter\newcommand\csname bb#1\endcsname{{\mathbb{#1}}}
  \expandafter\newcommand\csname cal#1\endcsname{{\mathcal{#1}}}   
  \expandafter\newcommand\csname rm#1\endcsname{{\mathrm{#1}}}
  \expandafter\newcommand\csname bf#1\endcsname{{\mathbf{#1}}}
  \expandafter\newcommand\csname bold#1\endcsname{{\boldsymbol{#1}}}
  \expandafter\newcommand\csname hat#1\endcsname{\hat{#1}}
  \expandafter\newcommand\csname tilde#1\endcsname{\widetilde{#1}}
  \expandafter\newcommand\csname bar#1\endcsname{\overline{#1}}
  \expandafter\newcommand\csname frak#1\endcsname{\mathfrak{#1}}
  }
\DoWith\UpperCase ABCDEFGHILJKMNOPQRSTUVWXYZ\StopDoing

\newcommand{\LowerCase}[1]{
  \expandafter\newcommand\csname rm#1\endcsname{{\mathrm{#1}}} 
  \expandafter\newcommand\csname bf#1\endcsname{{\mathbf{#1}}} 
  \expandafter\newcommand\csname bold#1\endcsname{{\boldsymbol{#1}}}
  \expandafter\newcommand\csname hat#1\endcsname{\hat{#1}}
  \expandafter\newcommand\csname tilde#1\endcsname{\tilde{#1}}
  \expandafter\newcommand\csname bar#1\endcsname{\bar{#1}}
  \expandafter\newcommand\csname frak#1\endcsname{\mathfrak{#1}}
  \expandafter\newcommand\csname vec#1\endcsname{\vec{#1}}
  }
\DoWith\LowerCase abcdefghijklmnopqrstuvwxyz\StopDoing

\newcommand{\MeasureUpperCase}[1]{
  \expandafter\newcommand\csname d#1\endcsname{\mathrm{d}#1}
  \expandafter\newcommand\csname dvec#1\endcsname{\mathrm{d}\vec{#1}}
  }
\DoWith\MeasureUpperCase ABCDEFGHILJKMNOPQRSTUVWXYZ\StopDoing

\newcommand{\MeasureLowerCase}[1]{
  \expandafter\newcommand\csname d#1\endcsname{\mathrm{d}#1}
  \expandafter\newcommand\csname dvec#1\endcsname{\mathrm{d}\vec{#1}}
  }
\DoWith\MeasureLowerCase abcegiklmnrstuwxyz\StopDoing

\newcommand{\SU}[1]{{\rm SU}(#1)}

\newcommand{\MSbar}{\overline{\rm MS}}

\newcommand{\mg}{m_\rmg}
\newcommand{\mh}{m_\rmH}

\newcommand{\mt}{m_\rmt}

\newcommand{\xig}{\xi}

\newcommand{\barpsi}{\overline{\psi}}

\newcommand{\ghost}{{\rm Ghost}}
\newcommand{\GF}{{\rm GF}}

\newcommand{\eq}{Eq.~}

\newcommand{\ep}{\epsilon}

\newcommand{\fn}{{\rm fn}}

\newcommand{\Sec}{Section~}

\newcommand{\TR}{T_\rmR}
\newcommand{\nf}{{n_\rmf}}

\newcommand{\Ca}{C_\rmA}
\newcommand{\Cf}{C_\rmF}

\newcommand{\as}{\alpha_{\rms}}

\newcommand{\OS}{\rm OS}
\newcommand{\OSS}{\rm OS}
\newcommand{\MS}{\rm \overline{MS}}
\newcommand{\psit}{\psi_{\rm t}}

\usepackage[normalem]{ulem}

\usepackage{dashrule}

\newcommand{\OO}{\mathcal{O}}
\newcommand{\ali}[1]{\begin{align}#1\end{align}}
\newcommand{\bs}{\begin{subequations}}
\newcommand{\es}{\end{subequations}}
\def\fn{\footnote}
\definecolor{dgreen}{rgb}{0,0.2,0} 
\definecolor{mille}{rgb}{0.67,0.608,0.388} 
\definecolor{sbrown}{RGB}{139,69,19} 
\definecolor{myblueold}{RGB}{65,105,225}
\def\bd{\begin{dmath}} 
\def\ed{\end{dmath}}

\newcounter{DFQ}

\allowdisplaybreaks 
\begin{document}

\maketitle
\flushbottom


\newcommand{\FigTwoPointBlob}[3]{
    \centering
    \begin{tikzpicture}[baseline={([yshift=0.3mm]c.base)}]
    \begin{feynman}
        \vertex [blob, fill=black!10] (c) {$#1$};
        \vertex [left=1cm of c] (i);
        \vertex [right=1cm of c] (f);
        \diagram{
            (c) -- [#2] (i);
            (c) -- [#3] (f);
        };
    \end{feynman}
    \end{tikzpicture}
}

\newcommand{\FigTwoPointA}{
    \centering
    \begin{tikzpicture}[baseline={([yshift=-1mm]c.base)}]
    \begin{feynman}
        \vertex (c);
        \vertex [left=0.9cm of c] (i1);
        \vertex [right=0.9cm of c] (f1);
        \vertex [left=0.3cm of i1] (i2);
        \vertex [right=0.3cm of f1] (f2);
        \diagram{ 
            (i2) -- [plain] (i1) -- [fermion] (f1) -- [plain] (f2);  
        };
        \draw [gluon] (f1) arc[start angle=0, end angle=180, radius=0.9cm];
        \vertex [blob, fill=black!10, above=0.5cm of c] (c2){$\rm 1L$};
        \vertex [below=0.5cm of c, anchor=center, inner sep=0pt] (A) {(A)};
    \end{feynman}
    \end{tikzpicture}
}

\newcommand{\FigTwoPointB}{
    \centering
    \begin{tikzpicture}[baseline={([yshift=-1mm]c.base)}]
    \begin{feynman}
        \vertex (c);
        \vertex [left=0.9cm of c] (i1);
        \vertex [right=0.9cm of c] (f1);
        \vertex [left=0.3cm of i1] (i2);
        \vertex [right=0.3cm of f1] (f2);
        \vertex [above=0.9cm of c] (c2) ;
        \diagram{ 
            (i2) -- [plain] (i1) -- [fermion] (c) -- [fermion] (f1) -- [plain] (f2); 
            (c2) -- [gluon] (c);
        };
        \draw [gluon] (f1) arc[start angle=0, end angle=90, radius=0.9cm];
        \draw [gluon] (c2) arc[start angle=90, end angle=180, radius=0.9cm];
        \vertex [below=0.5cm of c, anchor=center, inner sep=0pt] (B) {(B)};
    \end{feynman}
    \end{tikzpicture}
}

\newcommand{\FigTwoPointC}{
    \centering
    \begin{tikzpicture}[baseline={([yshift=-1mm]c.base)}]
    \begin{feynman}
        \vertex (c);
        \vertex [right=0.9cm of c] (c2);
        \vertex [left=0.9cm of c] (i1);
        \vertex [left=0.3cm of i1] (i2);
        \vertex [right=0.9cm of c2] (f1);
        \vertex [right=0.3cm of f1] (f2);
        \diagram{ 
            (i2) -- [plain] (i1) -- [fermion] (c) -- [fermion] (c2) -- [fermion] (f1) -- [plain] (f2);   
        };
        \draw [gluon] (f1) arc[start angle=0, end angle=180, radius=0.9cm];
        \draw[fill=white, draw=white] (0.41,0.77) circle (0.15cm);
        \draw [gluon] (c2) arc[start angle=0, end angle=180, radius=0.9cm];
        \vertex [below right=0.5cm and 0.5cm of c, anchor=center, inner sep=0pt] (C) {(C)};
    \end{feynman}
    \end{tikzpicture}
}

\newcommand{\FigTwoPointD}{
    \centering
    \begin{tikzpicture}[baseline={([yshift=-1mm]c.base)}]
    \begin{feynman}
        \vertex (c);
        \vertex [left=0.45cm of c] (c2);
        \vertex [right=0.45cm of c] (c3);
        \vertex [left=0.45cm of c2] (i1);
        \vertex [left=0.3cm of i1] (i2);
        \vertex [right=0.45cm of c3] (f1);
        \vertex [right=0.3cm of f1] (f2);
        \diagram{ 
            (i2) -- [plain] (i1) -- [plain] (c2) -- [fermion] (c3) -- [plain] (f1) -- [plain] (f2);  
        };
        \draw [gluon] (f1) arc[start angle=0, end angle=180, radius=0.9cm];
        \draw [gluon] (0.45,0) arc[start angle=0, end angle=180, radius=0.45cm];
        \vertex [below=0.5cm of c, anchor=center, inner sep=0pt] (D) {(D)};
    \end{feynman}
    \end{tikzpicture}
}

\newcommand{\FigOneLoopRenA}{
    \centering
    \begin{tikzpicture}[baseline={([yshift=-1mm]c.base)}]
    \begin{feynman}
        \vertex (c);
        \vertex [left=0.9cm of c] (i1);
        \vertex [right=0.9cm of c] (f1);
        \vertex [left=0.3cm of i1] (i2);
        \vertex [right=0.3cm of f1] (f2);
        \diagram{ 
            (i2) -- [plain] (i1) -- [fermion] (f1) -- [plain] (f2); 
        };
        \draw [gluon] (f1) arc[start angle=0, end angle=180, radius=0.9cm];
       \vertex [star, star points=5, star point ratio=2.5, scale=0.46, fill=black, draw=black, left=0.9cm of c, anchor=center] (c2) {};
       \vertex [below=0.5cm of c, anchor=center, inner sep=0pt] (A) {(a)};
    \end{feynman}
    \end{tikzpicture}
}

\newcommand{\FigOneLoopRenB}{
    \centering
    \begin{tikzpicture}[baseline={([yshift=-1mm]c.base)}]
    \begin{feynman}
        \vertex (c);
        \vertex [left=0.9cm of c] (i1);
        \vertex [right=0.9cm of c] (f1);
        \vertex [left=0.3cm of i1] (i2);
        \vertex [right=0.3cm of f1] (f2);
        \diagram{ 
            (i2) -- [plain] (i1) -- [fermion] (f1) -- [plain] (f2); 
        };
        \draw [gluon] (f1) arc[start angle=0, end angle=180, radius=0.9cm];
        \vertex [star, star points=5, star point ratio=2.5, scale=0.46, fill=black, draw=black, right=0.96cm of c, anchor=center] (c2) {};
       \vertex [below=0.5cm of c, anchor=center, inner sep=0pt] (B) {(b)};
    \end{feynman}
    \end{tikzpicture}
}

\newcommand{\FigOneLoopRenC}{
    \centering
    \begin{tikzpicture}[baseline={([yshift=-1mm]c.base)}]
    \begin{feynman}
        \vertex (c);
        \vertex [left=0.9cm of c] (i1);
        \vertex [right=0.9cm of c] (f1);
        \vertex [left=0.3cm of i1] (i2);
        \vertex [right=0.3cm of f1] (f2);
        \diagram{ 
            (i2) -- [plain] (i1) -- [fermion] (f1) -- [plain] (f2); 
        };
        \draw [gluon] (f1) arc[start angle=0, end angle=180, radius=0.9cm];
        \vertex [star, star points=5, star point ratio=2.5, scale=0.46, fill=black, draw=black, above=0.845cm of c] (c2) {};
       \vertex [below=0.5cm of c, anchor=center, inner sep=0pt] (C) {(c)};
    \end{feynman}
    \end{tikzpicture}
}

\newcommand{\FigOneLoopRenD}{
    \centering
    \begin{tikzpicture}[baseline={([yshift=-1mm]c.base)}]
    \begin{feynman}
        \vertex [star, star points=5, star point ratio=2.5, scale=0.46, fill=black, draw=black] (c) {};
        \vertex [left=0.9cm of c] (i1);
        \vertex [right=0.9cm of c] (f1);
        \vertex [left=0.3cm of i1] (i2);
        \vertex [right=0.3cm of f1] (f2);
        \diagram{ 
            (i2) -- [plain] (f2); 
        };
        \draw [gluon] (f1) arc[start angle=0, end angle=180, radius=0.9cm];
       \vertex [below=0.5cm of c, anchor=center, inner sep=0pt] (D) {(d)};
    \end{feynman}
    \end{tikzpicture}
}

\newcommand{\FigOneLoopRenE}{
    \centering
    \begin{tikzpicture}[baseline={([yshift=-1mm]c.base)}]
    \begin{feynman}
        \vertex (c); 
        \vertex [left=0.7cm of c] (i1);
        \vertex [right=0.7cm of c] (f1);
        \diagram{ 
            (i1) -- [plain] (f1); 
        };
        \vertex at (c) [star, star points=5, star point ratio=2.5, scale=0.46, fill=white, draw=black] (star) {};
       \vertex [below=0.5cm of c, anchor=center, inner sep=0pt] (E) {(e)};
    \end{feynman}
    \end{tikzpicture}
}
\newcommand{\FigTwoPointBlobWithLabel}[4]{
    \centering
    \begin{tikzpicture}[baseline={([yshift=0.3mm]c.base)}]
    \begin{feynman}
        \vertex [blob, fill=black!10] (c) {$#1$};
        \vertex [left=1cm of c] (i);
        \vertex [right=1cm of c] (f);
        \diagram{
            (c) -- [#2] (i);
            (c) -- [#3] (f);
        };
        \vertex [below=0.8cm of c, anchor=center, inner sep=0pt] (v) {#4};
    \end{feynman}
    \end{tikzpicture}
}

\newcommand{\FigTwoPointBlobWithLabelQ}[3]{
    \centering
    \begin{tikzpicture}[baseline={([yshift=-0.5mm]c.base)}]
    \begin{feynman}
        \vertex [blob, fill=black!10] (c) {$q$};
        \vertex [left=1cm of c] (i);
        \vertex [right=1cm of c] (f);
        \diagram{
            (c) -- [#1] (i);
            (c) -- [#2] (f);
        };
        \vertex [below=0.8cm of c, anchor=center, inner sep=0pt] (v) {#3};
    \end{feynman}
    \end{tikzpicture}
}

\newcommand{\FigTwoPointBlobTadpoleWithLabel}[1]{
    \centering
    \begin{tikzpicture}[baseline={([yshift=-0.9mm]c.base)}]
    \begin{feynman}
        \vertex (c);
        \vertex [left=1cm of c] (i);
        \vertex [right=1cm of c] (f);
        \vertex [blob, fill=black!10, above=0.5cm of c] (c2) {$\rm tad$};
        \diagram{
            (c) -- [gluon] (i);
            (c) -- [gluon] (f);
            (c) -- [scalar] (c2);
        };
        \vertex [below=0.8cm of c, anchor=center, inner sep=0pt] (v) {#1};
    \end{feynman}
    \end{tikzpicture}
}

\newcommand{\FigTwoPointBlobQ}[2]{
    \centering
    \begin{tikzpicture}[baseline={([yshift=-0.5mm]c.base)}]
    \begin{feynman}
        \vertex [blob, fill=black!10] (c) {$q$};
        \vertex [left=1cm of c] (i);
        \vertex [right=1cm of c] (f);
        \diagram{
            (c) -- [#1] (i);
            (c) -- [#2] (f);
        };
    \end{feynman}
    \end{tikzpicture}
}

\newcommand{\FigTwoPointOneLoop}[4]{
    \centering
    \begin{tikzpicture}[baseline={([yshift=-1mm]c.base)}]
    \begin{feynman}
        \vertex (c);
        \vertex [right=0.9cm of c] (c2);
        \vertex [left=0.7cm of c] (i);
        \vertex [right=0.7cm of c2] (f);
        \diagram{
            (c) -- [gluon] (i); 
            (c2) -- [gluon] (f);
        };
        \draw [#1] (c2) arc[start angle=0, end angle=180, radius=0.45cm]
        node[midway, above] {$#3$};
        \draw [#2] (c) arc[start angle=180, end angle=360, radius=0.45cm]
        node[midway, below] {$#4$};
    \end{feynman}
    \end{tikzpicture}
}

\newcommand{\FigTwoPointBlobTadpole}{
    \centering
    \begin{tikzpicture}[baseline={([yshift=-0.9mm]c.base)}]
    \begin{feynman}
        \vertex (c);
        \vertex [left=1cm of c] (i);
        \vertex [right=1cm of c] (f);
        \vertex [blob, fill=black!10, above=0.5cm of c] (c2) {$\rm tad$};
        \diagram{
            (c) -- [gluon] (i);
            (c) -- [gluon] (f);
            (c) -- [scalar] (c2);
        };
    \end{feynman}
    \end{tikzpicture}
}

\newcommand{\FigTwoPointSimpleTadpole}[2]{
    \centering
    \begin{tikzpicture}[baseline={([yshift=-0.9mm]c.base)}]
    \begin{feynman}
        \vertex[dot, minimum size=0.1pt, inner sep=0pt] (c) {};
        \vertex [left=0.9cm of c] (i);
        \vertex [right=0.9cm of c] (f);
        \diagram{
            (c) -- [gluon] (i);
            (c) -- [gluon] (f);
        };
        \draw [#1] (c) to[out=45, in=135, looseness=2200] node[pos=0.20, right] {$#2$} (c);
    \end{feynman}
    \end{tikzpicture}
}

\newcommand{\FigTwoPointBalloonTadpole}[2]{
    \centering
    \begin{tikzpicture}[baseline={([yshift=-0.9mm]c.base)}]
    \begin{feynman}
        \vertex (c);
        \vertex [above=0.5cm of c] (c2) {};
        \vertex [left=1cm of c] (i);
        \vertex [right=1cm of c] (f);
        \diagram{
            (c) -- [gluon] (i);
            (c) -- [gluon] (f);
            (c) -- [scalar] (c2);
        };
        \draw [#1] (c2) arc[start angle=-90, end angle=270, radius=0.4cm] node[pos=0.25, right] {$#2$};
    \end{feynman}
    \end{tikzpicture}
}

\newcommand{\FigPropagator}[4]{
    \scalebox{#1}{
    \centering
    \begin{tikzpicture}[baseline={([yshift=-1mm]c.base)}]
    \begin{feynman}
        \vertex (a1) [label={[left=0.0cm] {$#2$}}];
        \vertex [right=1.15cm of a1] (c) [label={[right=0.0cm] {$#3$}}];
        \diagram{
            (a1) -- [#4, momentum={$p$}] (c);
        };
    \end{feynman} 
    \end{tikzpicture}
    }
}

\newcommand{\FigPropagatorWithStart}[5]{
    \scalebox{#1}{
    \centering
    \begin{tikzpicture}[baseline={([yshift=-1mm]c.base)}]
    \begin{feynman}
        \vertex (i) [label={[left=0.0cm] {$#2$}}];
        \vertex [right=1.4cm of i] (f) [label={[right=0.0cm] {$#3$}}];
        \vertex [right=0.7cm of i] (c);
        \diagram{
            (i) -- [#4, momentum={$p$}] (f);
        };
        \vertex [star, star points=5, star point ratio=2.5, scale=0.52, fill=#5, draw=black, above=0.02cm of c, anchor=center] (star) {};
    \end{feynman} 
    \end{tikzpicture}
    }
}

\newcommand{\FigThreePointVerex}[7]{
    \scalebox{#1}{
    \begin{tikzpicture}[baseline={([yshift=-1mm]c.base)}]
    \begin{feynman}
        \vertex (c);
        \vertex [above left=0.5cm and 1cm of c] (a1) [label={[left=0.0cm] {$#2$}}];
        \vertex [below left=0.5cm and 1cm of c] (a2) [label={[left=0.0cm] {$#3$}}];
        \vertex [right=1.1cm of c] (a3) [label={[right=0.0cm] {$#4$}}];
        \diagram{
            (c)  -- [#5] (a1);
            (a2) -- [#6] (c);
            (a3) -- [#7] (c);
        };
    \end{feynman} 
    \end{tikzpicture}
    }
}

\newcommand{\FigThreePointVerexWithStar}[7]{
    \scalebox{#1}{
    \begin{tikzpicture}[baseline={([yshift=-1mm]c.base)}]
    \begin{feynman}
        \vertex [star, star points=5, star point ratio=2.5, scale=0.52, fill=black, draw=black] (c) {};
        \vertex [above left=0.5cm and 1cm of c] (a1) [label={[left=0.0cm] {$#2$}}];
        \vertex [below left=0.5cm and 1cm of c] (a2) [label={[left=0.0cm] {$#3$}}];
        \vertex [right=1.1cm of c] (a3) [label={[right=0.0cm] {$#4$}}];
        \diagram{
            (c)  -- [#5] (a1);
            (a2) -- [#6] (c);
            (a3) -- [#7] (c);
        };
    \end{feynman} 
    \end{tikzpicture}
    }
}

\newcommand{\FigFourPointVerex}[9]{
    \scalebox{#1}{
        \centering
        \begin{tikzpicture}[baseline={([yshift=-1mm]c.base)}]
        \begin{feynman}
            \vertex (c);
            \vertex [above left=0.5cm and 1cm of c] (a1) [label={[left=0.0cm] {$#2$}}];
            \vertex [below left=0.5cm and 1cm of c] (a2) [label={[left=0.0cm] {$#3$}}];
            \vertex [below right=0.5cm and 1cm of c] (a3) [label={[right=0.0cm] {$#4$}}];
            \vertex [above right=0.5cm and 1cm of c] (a4) [label={[right=0.0cm] {$#5$}}];
            \diagram{
                (c)  -- [#6] (a1);
                (a2) -- [#7] (c);
                (c)  -- [#8] (a3);
                (a4) -- [#9] (c);
            };
        \end{feynman} 
        \end{tikzpicture}
    }
}


\section{Introduction}
\label{sec:introduction}

An ever-increasing precision of experiments in high-energy 
particle physics requires significant improvements 
of theoretical predictions. One aspect of such improvements involves non-perturbative power-suppressed effects. The importance of these effects for a particular process or observable is 
determined by the ratio $\Lambda_{\rm QCD}/Q$, with $\Lambda_{
\rm QCD}$ being the non-perturbative energy scale of Quantum Chromodynamics (QCD) and $Q$ a typical hard scale of the process. 
Since at high-energy colliders $Q \gg \Lambda_{\rm QCD}$, these corrections can be relevant only when the \textit{first power} of this ratio appears in theoretical predictions, and only for observables known with extraordinary precision. 
Well-known examples of such observables are the mass of the $W$ boson \cite{ATLAS:2024erm}, the mass of the top quark determined from the production of top quark pairs \cite{ATLAS:2024dxp}, and the strong coupling constant $\as$ extracted either from event shapes in $e^+e^-$ annihilation \cite{Becher:2008cf,Bell:2023dqs,Nason:2023asn,Benitez:2024nav,Nason:2025qbx} or from the transverse momentum distribution of $Z$ bosons at the Large Hadron Collider (LHC) \cite{Camarda:2022qdg,ATLAS:2023lhg}. This list is expected to grow as more and more high-precision measurements of various physical quantities will be performed at the high-luminosity LHC and also at future colliders. 

It is therefore interesting to understand how power corrections arise in the hard-scattering collider processes underlying precision measurements. 
However, this task is difficult as the conventional operator product expansion 
\cite{Wilson:1969zs,Shifman:1978bx} cannot be used to describe non-perturbative effects in hard partonic collisions. In this situation, the best one can do is to rely on models, either based 
on phenomenological descriptions of hadronization as implemented in parton-shower event generators
\cite{Andersson:1983ia,Sjostrand:1982fn,Sjostrand:1986hx,Webber:1983if}, 
or on analytic methods such as renormalon-based approaches \cite{Gross:1974jv,Lautrup:1977hs,tHooft:1977xjm,Parisi:1978bj,Mueller:1984vh}. The latter proceed by investigating the sensitivity of observables to the Landau pole in the running of the strong coupling constant. Technically, such sensitivity can be studied by computing perturbative QCD corrections with massive gluons \cite{Webber:1994cp,Beneke:1994bc,Beneke:1995pq,Ball:1995ni,Dokshitzer:1995qm}. This framework assumes that the mass of the gluon is the smallest parameter in the problem; then, by studying the dependence of the result on the gluon mass, it uses known formulas to translate powers of the gluon mass 
to powers of $\Lambda_{
\rm QCD}$ (for a review, see Ref.~\cite{Beneke:1998ui}).

Unfortunately, giving a mass to a gluon  without further modifications of the theory can only work in the Abelianized version of QCD at the lowest perturbative order. For collider processes, this implies that only processes with no gluons at tree level can be studied. This limitation was one of the factors that significantly slowed the application of such methods to collider processes. In fact, after early studies~\cite{Beneke:1994sw,Bigi:1994em,Manohar:1994kq,Webber:1994cp,Korchemsky:1994is,Beneke:1995pq,Akhoury:1995sp,Dokshitzer:1995zt,Nason:1995np,Dokshitzer:1995qm,Dasgupta:1996ki,Korchemsky:1996iq,Beneke:1997sr,Dokshitzer:1997ew,Dokshitzer:1997iz,Dokshitzer:1998pt,Dasgupta:1999zm}, the subject saw relatively little development for almost two decades, apart from an investigation of linear power corrections to jet observables~\cite{Dasgupta:2007wa}. Systematic attempts to extend such analyses to more complex processes -- serving as proxies for realistic reactions at hadron colliders -- were only undertaken much later~\cite{FerrarioRavasio:2018ubr,FerrarioRavasio:2020guj,FerrarioRavasio:2021mzg}. Even now, computations reported in Refs \cite{Caola:2021kzt,Makarov:2023ttq} in the context of renormalon models deal with LHC processes without gluons in Born processes. 
If this is insufficient for phenomenology, 
one adjusts the particle content of the process 
of interest (for example, by changing gluons to photons), and then argues that the results derived in the simplified cases are likely to hold also in the real ones (see e.g.\ Refs \cite{Caola:2021kzt,Caola:2022vea}).

It is interesting to ask if one can do better than that. The question is especially motivated by arguments suggesting that linear corrections must exhibit a much simpler structure, including their complete cancellation in inclusive observables and rates \cite{Akhoury:1995cj}. 
This reasoning can be viewed as a generalization of the Bloch–Nordsieck \cite{Bloch:1937pw} and Kinoshita–Lee–Nauenberg \cite{Kinoshita:1962ur,Lee:1964is} 
theorems beyond the logarithmic accuracy. While the cancellation of linear power corrections has been explicitly demonstrated in the Abelianized version of QCD for relatively simple processes \cite{Beneke:1994bc,Akhoury:1996ks,Makarov:2023ttq,Makarov:2023uet}, an unambiguous proof in the non-Abelian case -- as well as a framework 
to allow their calculation for realistic collider observables in QCD -- 
is still lacking (see, however, Ref.~\cite{Sinkovics:1998mi}). This question is of direct phenomenological relevance, since hard-scattering processes at colliders are often governed by QCD and, hence, involve non-Abelian dynamics.

A useful hint on how to proceed comes again from the renormalon picture, where the problem of non-perturbative sensitivity is effectively mapped onto the problem of understanding radiative corrections with a massive gluon. This suggests promoting the gluon mass from a purely technical device to a parameter of a consistent quantum field theory, where it quantifies the degree of infrared (IR) sensitivity. A natural realization of this idea is provided by a non-Abelian gauge theory in which the gauge symmetry is spontaneously broken via the Higgs mechanism. 
The gluon then acquires a small mass $\mg$ in a theoretically consistent way, providing a framework in which the dependence of IR-safe observables on this mass can be systematically analyzed. In particular, a \textit{linear} dependence of an observable on $\mg$ is a signal of \textit{linear} IR sensitivity.

In this paper, we construct such a theory and discuss it in detail. Besides containing a heavy quark and a gluon, it requires the usual ingredients of a properly-quantized non-Abelian gauge theory with spontaneous symmetry breaking, including gauge fixing, ghost fields and the scalar degrees of freedom associated with the symmetry-breaking sector. This toy model provides a useful laboratory for studying the IR sensitivity of collider processes, in particular the production and decay of heavy quarks.

As a first step towards exploring such processes, we calculate two ingredients needed for their study: the $\order{\mg}$ contributions to 
the relations between the pole and the modified minimal subtraction $(\overline{\rm MS})$ masses of a heavy quark and between the corresponding field counterterms in the on-shell subtraction (OS) and $\overline{\rm MS}$ schemes. 
Both relations are calculated through two loops, as this is the first perturbative order in which genuinely non-Abelian dynamics arises.

The relation between masses is a classic example of an observable with linear IR sensitivity \cite{Bigi:1994em,Beneke:1994sw,Beneke:1994rs}.\footnote{These papers triggered many studies of mass-related linear sensitivity to non-perturbative physics in various physical systems and, among other things, led to definitions of the so-called short-distance low-scale masses \cite{Hoang:2008yj,Beneke:1998rk,Bigi:1997fj,Hoang:1998hm,Hoang:2018zrp}.
}
It is instructive to derive  it because the calculation is nontrivial, yet sufficiently simple to be discussed in detail. We therefore analyze it carefully and show that it contains a term linear in $\mg$, reflecting the expected IR sensitivity. Our analysis clarifies the origin of this linear correction and explains how analytic computations of terms linear in $\mg$ can be performed. 

We emphasize that developing a computational framework which can be applied to compute $\order{\mg}$ effects in collider observables is important, especially since their analysis is technically more demanding than the mass relation considered here. We demonstrate that the computational framework described below is flexible, as it also allows us to study the relation between heavy-quark field counterterms, which exhibits both logarithmic and linear sensitivities to the gluon mass. 

Finally, we note that there is a direct connection between $\order{\mg}$ contributions to the relation between the 
$\overline{\rm MS}$ and pole masses of a heavy quark, and the mass splittings in supersymmetric multiplets induced by loops of electroweak gauge bosons. It is known that such splittings at one loop are proportional to the \emph{first power} of the $W$-boson mass \cite{Mizuta:1992ja,Gherghetta:1999sw}.\footnote{We  thank  X.~Tata for pointing this out to us.}
A similar result was obtained numerically in a slightly more general context at the two-loop level in Ref.~\cite{Yamada:2009ve}. 
Our discussion elucidates the origin of these corrections and suggests a way to compute them 
systematically in higher orders of perturbation theory. 

The remainder of the paper is organized as follows. 
We start by describing the toy model in \Sec\ref{sec:model}. We  devote \Sec\ref{sec:calculation} to the two-loop calculations of $\OO(\mg)$ contributions to the  relations  between the pole and the 
$\overline{\rm MS}$ mass of the heavy quark, and the relation between the heavy-quark field counterterm in the $\OSS$ and $\overline{\rm MS}$ schemes. 
We present conclusions in \Sec\ref{sec:conclusions}. 
Additional technical details can be found in appendices.

\section{The model}
\label{sec:model}

The toy model employed for the computations presented in this paper is a renormalizable non-Abelian gauge theory where the gauge group ${\rm SU}(2)$ is spontaneously broken by the Higgs mechanism. 
Inspired by QCD, we refer to the gauge fields as gluons and to the massive fermion as top quark. To make the model QCD-like, we also include $\nf$ flavours of massless quarks. 
Since our goal is to perform perturbative computations in this model through two loops, we will have to discuss its renormalization. As usual, we assume dimensional regularization with $d=4-2 \ep$ dimensions and a renormalization scale $\mu$.

In the remainder of this section, we describe the Lagrangian and the particle content of the theory. We then continue with a brief discussion of the renormalization. 

\subsection{The Lagrangian of the model and its particle content}

The Lagrangian of the model reads%
\begin{equation}
\begin{split}
 \calL
 = & - \frac{1}{4} G^{a}_{\mu\nu} G^{a,\mu \nu}
 + \barpsi_\rmt (\rmi\slashed{D} - \mt) \psi_\rmt 
 + \sum \limits_{k=1}^{\nf} 
 \barpsi_k \rmi \slashed{D} \psi_k \\
 & + (D_\mu \Phi)^\dagger (D^\mu \Phi) 
 + \mu^2 \Phi^\dagger \Phi 
 - \lambda (\Phi^\dagger \Phi)^2
 + \calL_\GF + \calL_\ghost \,.
\label{eq_Lagrangian}
\end{split}
\end{equation}
Here, $\mu^2 > 0 $ and $\lambda > 0$ are real parameters, $\Phi$ is the Higgs doublet,
$\psi_\rmt$ and $\mt$ are the top-quark field and mass, respectively, and $\psi_{k= 1,...,\nf}$ represent fields of $\nf$ massless quarks.
Furthermore, $\calL_\GF$ and $\calL_\ghost$ are the gauge-fixing and ghost terms, respectively. 
The quantities $D_\mu$ and $G_{\mu\nu}^a$ are the covariant derivative and the field-strength tensor. They are  defined in the standard way as
\be
D_\mu = \partial_\mu + \rmi g \hatT^a G^a_{\mu}, 
\qquad
G_{\mu\nu}^a = \partial_\mu G_\nu^a - \partial_\nu G_\mu^a - g f^{abc} G_\mu^b G_\nu^c \,,
\ee
where $G_{\mu}^{a}$ is the gluon field, $g$ is the strong coupling constant, and 
$\hatT^a$ and $f^{abc}$ denote generators and structure constants of the gauge group.
For the $\SU{2}$ case, one has $f^{abc} = \ep^{abc}$ and $\hatT^a = \sigma^a/2$, where $\sigma^a$ are the Pauli matrices. We will employ generic Casimir 
operators $\Ca$, $\Cf$ and $\TR$ in what follows, to distinguish 
the origin of different terms in the results. 

Since $\mu^2 > 0$, 
the Higgs field acquires a non-vanishing expectation value, breaking the gauge symmetry spontaneously. To make this explicit, we parametrize the Higgs doublet as 
\begin{equation}
 \Phi(x) = \frac{1}{\sqrt{2}}
 \begin{pmatrix}
 - \varphi_2(x) - \rmi\varphi_1(x) \\ v + H(x) + \rmi\varphi_3(x)
 \end{pmatrix} \,,
\label{eq_Phi_doublet_def}
\end{equation}
where $\varphi_{1,2,3}$ are the would-be Goldstone bosons, $H$ is the physical Higgs field and $v$ is the vacuum expectation value.
The gluon and the Higgs boson acquire masses $\mg$ and $\mh$, proportional to the vacuum expectation value; they are given by
\begin{equation}
    \label{eq:parameter-relations}
    \mg = \frac{gv}{2} \,, 
    \qquad
    \mh = \sqrt{2 \lambda} \, v \,,
    \quad {\rm with} \quad
    v = \sqrt{\frac{\mu^2}{\lambda}} \,.
\end{equation}
Finally, the gauge-fixing term
$\calL_\GF$ and the ghost Lagrangian $\calL_\ghost $
read 
\begin{equation}
\begin{split}
    & \calL_\GF = - \frac{1}{2\xig} F_a F_a \,,
    \hspace{9mm}
    F_a = \partial^{\mu} G_{\mu}^{a} + \xig \mg \varphi_a, \quad a=1,2,3 \,, \\
    & 
    \begin{aligned}
        \calL_\ghost 
        =&  - \barc_a(\partial_\mu \partial^\mu + \xig \mg^2) c_a
        - g \ep^{abc} \barc_a \Big[(\partial_\mu c_b) G_c^\mu + c_b (\partial_\mu G_c^\mu) \Big] \\
        & - \frac{g \xig \mg}{2} \barc_a c_a H
        - \frac{g \xig \mg}{2} \ep^{abc} \barc_a c_b \varphi_c \,,
    \end{aligned}
 \end{split}
\end{equation}
where $\xi$ is the gauge parameter, and $c_a$ and $\bar{c}_a$ are the ghost and the anti-ghost field. 

The particle content of the theory follows from the above Lagrangian
in a standard manner. 
After the symmetry breaking, the spectrum consists of the top quark of mass $\mt$, $\nf$ massless quarks, three massive gluons of mass $\mg$, and a single Higgs boson of mass $\mh$.%
\footnote{The top quark mass is an original parameter of this theory. In fact, at variance with the $\SU{2}$ gauge sector of the Glashow–Weinberg–Salam model \cite{Glashow:1961tr, Salam:1968rm, Weinberg:1967tq}, the present theory is not chiral. Therefore,  a Dirac mass term for the top quark is included from the outset. Interestingly, the interaction between the Higgs boson and the top quark is loop induced in this setup.}
When working in a non-unitary gauge, the would-be Goldstone bosons also appear in the spectrum of the theory. 
Finally, as it is common to all non-Abelian gauge theories, unphysical ghost and anti-ghost fields are also present.


\subsection{Renormalization}

We use \eq\eqref{eq:parameter-relations} to select $\mh, \mg, \mt$ and $g$ as independent physical quantities. When these parameters appear in the original Lagrangian, they should be considered as bare and should be written in terms of the renormalized ones,
\begin{equation}
    \label{eq:parameter-CTs}
    m_{{\rm H}(0)}^2 = Z_{\mh^2} \mh^2 \,,
    \quad
    m_{{\rm g}(0)}^2 = Z_{\mg^2} \mg^2 \,,
    \quad
    m_{{\rm t}(0)} = Z_{\mt} \mt \,, 
    \quad
    g_{(0)} = \big[S_\ep \mu^{2\ep}\big]^{\frac12} Z_\rmg \, g \,,
\end{equation}
where $S_\ep = (4 \pi)^{-\ep} e^{\ep\gamma_\rmE}$.
Here, bare quantities are denoted with the subscript $(0)$ and the renormalized ones without it, and the $Z$ factors are the renormalization constants.
In a similar way, we also renormalize the fields,%
\fn{
Field counterterms for the would-be Goldstone bosons and the ghosts do not need to be specified since we do not consider Green's functions that involve those fields.}
\begin{equation}
\label{eq:field-CTs}
    \psi_{\rmt (0)} = Z^{1/2}_{\psit} 
    \psi_\rmt \,,
    \qquad
    \psi_{k (0)} = Z^{1/2}_{\psi_k} 
    \psi_{k} \,,
    \qquad
    H_{(0)} = Z_\rmH^{1/2} H \,,
    \qquad
    G^{a}_{\mu(0)} = Z_\rmG^{1/2} G^{a}_{\mu} \,.
\end{equation}
While the $Z$ factors for both parameters and fields are a priori unknown, they are fixed through renormalization conditions discussed below.
For a generic parameter or field $X$, we define the counterterm $\delta Z_X$ such that $Z_X = 1 + \delta Z_X$. Furthermore, we write $\delta Z_X = \sum_i \delta Z_X^{(i)}$, with $\delta Z_X^{(i)}$ denoting the $i$-th loop contributions to $\delta Z_X$.

We note that, for the purpose of renormalization, the vacuum expectation value is   treated as a dependent bare parameter.  
This ensures that all mass counterterms are gauge independent. 
On the other hand, Higgs tadpole diagrams must be included in all possible Green's functions to which they can contribute. A detailed discussion of these topics can be found in Ref.~\cite{deSousaMachadoFontes:2021zia}. 

We now briefly discuss our renormalization scheme choices.
All ultraviolet (UV) counterterms listed above, with the exception of $\delta Z_\rmg$, are computed in the $\OSS$ scheme \cite{Ross:1973fp}. Explicit conditions that define this scheme, as well as calculations that enable the determination of the corresponding counterterms, are discussed e.g.~in Refs \cite{Denner:2019vbn,deSousaMachadoFontes:2021zia}.
To find $\delta Z_\rmg$, we first determine its divergent part. This can be done by considering the amplitude for $G\to t\bar t$ and requiring it to be UV-finite. The counterterm vertex for this process depends on $\delta Z_\rmg$, as well as the field counterterms $\delta Z_{\rm G}$ and $\delta Z_\rmt$. 
Since $\delta Z_\rmG$ and $\delta Z_\rmt$ are known from the $\OSS$  conditions, computing the UV-divergent part of the one-loop $G\to t\bar t$ amplitude allows us to obtain the UV-divergent part of $\delta Z_\rmg$.

The above procedure is sufficient to determine the 
coupling constant in the $\MS$ scheme.
Although the $\MS$ scheme  can be used to describe the ${\cal O}(\mg)$ contributions to the renormalization constants, 
it is more useful to define a coupling 
constant that absorbs certain finite parts of loop diagrams, in analogy with the renormalization of the fine-structure constant in QED. 
To this end, we additionally require 
that the coupling constant counterterm, $\delta Z_\rmg$, absorbs the finite contribution from top quark and Higgs boson contributions to the gluon-field counterterm $\delta Z_\rmG$. 
This definition is appropriate for renormalization scales below the top quark mass and the mass of the Higgs boson. 
We note that the running of the 
strong coupling constant $\as = g^2/(4\pi)$ defined in 
this way is controlled by the equation 
\begin{equation}
 \mu \dv{\as}{\mu} = -\frac{\beta_0}{2\pi} \as^2 \,,
 \qquad
 \beta_0 = \frac{29}{8}\Ca - \frac{4}{3} \TR \nf \,.
 \label{eq:beta-MS}
\end{equation}
Similarly to QCD, as long as the number of massless fermion species is not too high, the theory is asymptotically free. 
Finally, we note that Feynman rules for the renormalized interactions 
and relevant counterterms can be found in Appendix~\ref{App_Feynman_rules}.
\vspace{-0.5mm}

\section{Calculation}
\label{sec:calculation}

We now turn to the relation between the pole mass and the $\overline{\rm MS}$ mass of the top quark, and the relation between the top-quark field counterterms in the $\OSS$ and $\MSbar$ schemes. Our goal is to 
compute the ${\cal O}(\mg)$ 
contributions to these relations through two loops. Both relations can be obtained following the method of Ref.~\cite{Melnikov:2000zc}, which is formulated without explicit Feynman rules for the counterterms. That method can also be described in the framework of renormalized perturbation theory, which we use in the calculation described below. 
Details on the adaptation of the method of Ref.~\cite{Melnikov:2000zc} to that framework are given in Appendix~\ref{app:OS-conditions}. 
We note that we have performed the calculation using  both formulations, providing a useful cross-check of the results. 

The calculations are described 
in Sections~\ref{sec:preliminaries} and \ref{sec:masters}, and 
the results can be found in 
Sections~\ref{sec:relation-for-masses} and \ref{sec:relation-for-field}. We assume the hierarchy of masses $\mg \ll \mt \sim \mh$ and
focus on terms linear in $\mg$. 
In agreement with the conventions of Section \ref{sec:model}, $\mt$ (without superscripts) denotes the pole mass.

\subsection{Preliminaries}
\label{sec:preliminaries}

\begin{figure}[t!]
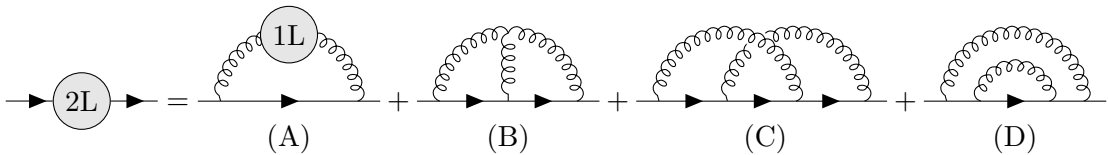

\centering
\begin{align*}
\FigTwoPointBlob{\rm 2L}{anti fermion}{fermion}
=
\FigTwoPointA + \FigTwoPointB + \FigTwoPointC + \FigTwoPointD
\end{align*}
\vspace{-1.3em}
\caption{Two-loop Feynman diagrams for the top-quark self-energy.}
\label{fig:bare}
\end{figure}

Using Eqs \eqref{eq:parameter-CTs} and \eqref{eq:field-CTs}, as well as the fact that the top quark pole mass is the mass   defined in the $\OSS$ scheme, we write two equalities for the bare quantities, 
\ali{
m_{{\rm t}(0)} = Z_{\mt}^{\rm OS} \mt = Z_{\mt}^{\MS}(\mu) \, \mt^{\MS}(\mu) \,,
\qquad
\psi_{{\rm t}(0)} = \sqrt{Z_{\psit}^{\OS}} \psit^{\OS} = \sqrt{Z_{\psit}^{\MS}} \psit^{\MS}(\mu) \,,
}
and use them to derive the
two relations that are of interest to us,
\ali{
\label{eq:the-relations}
 \frac{\mt^{{\rm \overline{MS}}}(\mu)}{\mt} = \frac{Z_{\mt}^{\rm OS}}{Z_{\mt}^{\MS}(\mu)} \,,
\qquad
\frac{\psit^{\MS}(\mu)}{\psit^{\OS}} = \sqrt{\frac{Z_{\psit}^{\OS}}{Z_{\psit}^{\MS}}} \,.
}
The renormalization constants 
are computed in perturbation theory. 
The two nontrivial quantities to be determined are the two-loop mass and field counterterms in the $\OSS$  scheme, $\delta Z_{\mt}^{\OS(2)}$ and $\delta Z_{\psit}^{\OS(2)}$. 
They are obtained from  the two-loop top-quark self-energy (cf.\ Appendix~\ref{app:OS-conditions}). The genuine two-loop Feynman diagrams are shown in Figure~\ref{fig:bare}.%
\fn{Details on the gluon one-loop vacuum polarization included in diagram (A) will be discussed below. For now, it is sufficient  to mention that it contains contributions from the top quark, the Higgs boson and gluons.}
To compute $\delta Z_{\mt}^{\OS(2)}$, it is sufficient to consider the top quark on shell and, hence, to evaluate the two-loop self-energy diagrams at an external momentum $p$ that satisfies 
$p^2=\mt^2$.
For $\delta Z_{\psit}^{\OS(2)}$, derivatives of the self-energy with respect to $p^2$ need to be computed before the on-shell limit $p^2=\mt^2$ can be taken.

The diagrams in Figure~\ref{fig:bare} depend on several mass scales, including $\mg$, $\mt$ and $\mh$. The presence of multiple scales makes  exact evaluation of these diagrams difficult. However, since we are only interested in contributions that are \emph{linear} in $\mg$, we can expand the integrands of the relevant Feynman diagrams in powers of $\mg$ \emph{before} performing the loop integrations. This is achieved by using the method known as ``strategy of regions'' \cite{Beneke:1997zp,Beneke:1999zr,Jantzen:2011nz}.%
\fn{It is worth noting that the expansion of one of the two-loop two-point integrals required here, in the limit of small gluon mass, was already discussed in Ref.~\cite{Czarnecki:1996nr}, providing an early example of such unconventional asymptotic expansions of Feynman diagrams.}

To explain how the expansion of the integrands is constructed, we consider a one-loop scalar integral contributing to the one-loop on-shell self-energy, 
\ali{
\label{eq:I-def}
I = \int \dfrac{\rmd^d k}{(2 \pi)^d} \dfrac{\mu^{2\ep}}{(k^2 - \mg^2)(k^2+2 p \cdot k)} \,,
\qquad
p^2 = \mt^2.
}
We start by calculating the integral exactly using standard methods. This is straightforward and, upon expanding the exact result in powers of $\mg/\mt$, we find
\ali{
\label{eq:I-expanded}
I = \dfrac{\rmi  \Gamma(1+\ep)}{(4 \pi)^{d/2}} \Bigg\{\left[ \dfrac{1}{\ep} + 2 \log \dfrac{\mu}{\mt} + 2 \right] - \pi \dfrac{\mg}{\mt} + \order{\mg^2} \Bigg\}.
}
We note that a term linear in the gluon mass $\mg$ is present in the above expression.

The strategy of regions allows us to obtain \eq\eqref{eq:I-expanded} by expanding \emph{the integrand} in \eq\eqref{eq:I-def} 
in small quantities. These quantities depend on 
the loop-momentum region. If the loop momentum is hard, $k \sim \mt$, the gluon mass $\mg$ is the only small parameter, so that one expands the gluon propagator in $\mg^2/k^2$. If, on the other hand, the loop momentum is soft, 
$k \sim \mg$, one expands the quark propagator in $k^2/(2k\cdot p)$. Performing the corresponding expansions to linear order in $\mg$, and denoting the expressions calculated in the hard and soft regions with superscripts $(\rmh)$ and $(\rms)$, respectively, we find
\begin{align}
\begin{aligned}
	\hspace{-2mm}I^{(\rmh)} & = \!\int \! \dfrac{\rmd^d k}{(2 \pi)^d } \dfrac{\mu^{2\ep}}{k^2 \left( k^2 - 2 p \cdot k\right)} + \order{\mg^2}
	= \dfrac{\rmi \Gamma(1+\ep) }{(4 \pi)^{d/2} } \bigg[\dfrac{1}{\ep} + 2 \log \dfrac{\mu}{\mt} + 2 \bigg] + \order{\mg^2} \,, \\
	\hspace{-2mm}I^{(\rms)} & = \! \int \! \dfrac{\rmd^d k}{(2 \pi)^d} \dfrac{\mu^{2\ep}}{\left(k^2 -\mg^2\right) \left( - 2 p \cdot k\right)} + \order{\mg^2} = 
    \dfrac{\rmi \Gamma(1+\ep)}{(4 \pi)^{d/2}} \left( - \pi \dfrac{\mg}{\mt} \right) + \order{\mg^2} \,.
 \label{eq3.2}
\end{aligned}
\end{align}
As expected, the result for the integral $I$ in \eq\eqref{eq:I-expanded} agrees with the sum of $I^{(\rmh)}$ and $I^{(\rms)}$ 
in\eq\eqref{eq3.2}.

The $\order{\mg}$ term is fully determined by the soft region, and it is certainly not surprising that the hard region cannot produce $\order{\mg}$ terms. Indeed, since the gluon propagator depends on $\mg^2$, and since 
the hard region corresponds to a Taylor expansion in  $\mg$, it can only generate even powers of $\mg$. By contrast, in the soft region $k \sim \mg$, and this power-counting immediately implies 
\be
I^{(\rms)} \sim \int \frac{\rmd^d k}{(2\pi)^d}
\; \frac{1}{k^2 (k \cdot p)} \stackrel{k \sim \mg}{\sim} \mg \,.
\ee

The approach described in this one-loop example extends to the two-loop case in a natural way. 
For a particular choice of the loop momenta $k_{1,2}$, four regions have to be considered:
\begin{equation}
\begin{aligned}
\label{eq.two-loop-regions}
 \text{hard-hard:} & ~ k_1 \sim \mt, ~ k_2 \sim \mt \,,
 \qquad & \text{hard-soft:} & ~ k_1 \sim \mt, \,\, k_2 \sim \mg \,, \\
 \text{soft-hard:} & ~ k_1 \sim \mg, ~ k_2 \sim \mt \,,
 \qquad & \text{soft-soft:} & ~ k_1 \sim \mg, \,\, k_2 \sim \mg \,.
\end{aligned} 
\end{equation}
Similarly to the one-loop case, only hard-soft, 
soft-hard and soft-soft regions can produce $\order{\mg}$ terms. 

\begin{figure}[t!]
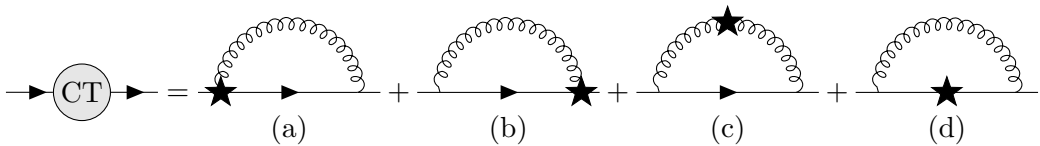

\centering
\begin{align*}
 \FigTwoPointBlob{\rm CT}{anti fermion}{fermion}
 = \FigOneLoopRenA + \FigOneLoopRenB + \FigOneLoopRenC + \FigOneLoopRenD
\end{align*}
\vspace{-1.3em}
\caption{Feynman diagrams for the top-quark self-energy with one-loop counterterm insertions.}
\label{fig:CTs}
\end{figure}

The calculation of the two nontrivial quantities of interest follows a standard workflow. 
First, we generate the total set of Feynman diagrams; besides those of Figure~\ref{fig:bare}, this includes the counterterm diagrams shown in Figure~\ref{fig:CTs}.
We then use the method of Appendix \ref{app:OS-conditions} to project all diagrams onto scalar integrals relevant for the computation of $\delta Z_{\mt}^{\OS(2)}$ and $\delta Z_{\psit}^{\OS(2)}$. Each scalar integral is expanded with the method of regions keeping only terms linear in $\mg$. The resulting integrals from the different regions are collected and reduced to a small set of master integrals using integration-by-parts identities~\cite{Chetyrkin:1981qh,Chetyrkin:1997dh}.
We have performed two independent implementations of the entire procedure,  from diagram generation and renormalization to integrand expansion and reduction to master integrals.  Several publicly available software packages were employed in this process, including FeynMaster \cite{Fontes:2019wqh,Fontes:2021iue,Fontes:2025svw}, FeynRules~\cite{Christensen:2008py,Alloul:2013bka}, Qgraf~\cite{Nogueira:1991ex}, FeynCalc~\cite{Mertig:1990an,Shtabovenko:2016sxi,Shtabovenko:2020gxv,Shtabovenko:2023idz,Shtabovenko:2025lxq}, FeynHelpers \cite{Shtabovenko:2016whf,Shtabovenko:2025lxq}, Package-X \cite{Patel:2015tea,Patel:2016fam}, LiteRed~\cite{Lee:2012cn,Lee:2013mka} and FIRE~\cite{Smirnov:2025prc}.

\subsection{Master integrals}
\label{sec:masters}

We now turn to the discussion of the master integrals.
The master integrals for the hard-soft and soft-hard regions defined  in \eq\eqref{eq.two-loop-regions} are simple, since the contributions of hard and soft loops factorize. 
Therefore, such integrals are written as products of one-loop on-shell integrals with $\mg = 0$ and the one-loop soft integral. 
We also find that only four double-soft integrals  are needed to compute 
${\cal O}(\mg)$
contributions to 
$\delta Z_{\mt}^{\OS(2)}$ and $\delta Z_{\psit}^{\OS(2)}$ 
terms. They are 
\begin{align}
    I_\rmA & = \int \frac{\rmd^d k_1}{(2\pi)^d} \frac{\rmd^d k_2}{(2\pi)^d} \, \frac{1}{(k_1^2 - \mg^2)} \frac{1}{(k_2^2 - \mg^2)} \frac{1}{(2k_2\cdot p)} \,, 
    \label{eq_I7_def} \\
    I_\rmB & = \int \frac{\rmd^d k_1}{(2\pi)^d} \frac{\rmd^d k_2}{(2\pi)^d} \, \frac{1}{(k_1^2 - \mg^2)} 
    \frac{1}{[(k_1-k_2)^2 - \mg^2]}
    \frac{1}{[2 k_2 \cdot p]} \,,
    \label{eq_I6_def} \\
    I_\rmC & = \int \frac{\rmd^d k_1}{(2\pi)^d} \frac{\rmd^d k_2}{(2\pi)^d} \, 
    \frac{1}{(k_1^2 - \mg^2)}
    \frac{1}{[(k_1 - k_2)^2 - \mg^2]}
    \frac{1}{(k_2^2 - \mg^2)}
    \frac{1}{(2k_2\cdot p)} \,, 
    \label{eq_I2_def} \\
    I_\rmD & = \int \frac{\rmd^d k_1}{(2\pi)^d} \frac{\rmd^d k_2}{(2\pi)^d} \, 
    \frac{1}{k_1^2}
    \frac{1}{(k_1 - k_2)^2}
    \frac{1}{(k_2^2 - \mg^2)} \frac{1}{(2k_2\cdot p)} \,,
    \label{eq_I8_def}
\end{align}
where $p^2 = \mt^2$, and the standard $+\rmi0$ prescription for each propagator is understood. 

The integral $I_\rmA$ is elementary, since  integrations over loop momenta $k_{1,2}$ factorize. 
For further reference, we write the one-loop integral explicitly,
\begin{equation}
    \int \frac{\rmd^d k_2}{(2\pi)^d} \, \frac{1}{(k_2^2 - \mg^2)} \frac{1}{(2k_2\cdot p)}
    = \frac{\mg^{1-2\ep}}{\mt} \frac{\rmi}{2(4 \pi)^{d/2}} \Gamma(1/2) \Gamma(-1/2+\ep) \,.
\label{eq_F_of_s_def}
\end{equation}
Using the known result for the massive tadpole, 
we find 
\begin{equation}
    I_\rmA = - \frac{\mg^{3-4\ep}}{\mt} \frac{1}{(4\pi)^{d}}
    \frac{\Gamma(1/2) \Gamma(1+\ep) \Gamma(-1/2+\ep)}{2\ep (1-\ep)} \,.
\end{equation}

The calculation of the three remaining integrals, $I_{\rmB,\rmC,\rmD}$, is slightly more involved. In principle, they can be evaluated using standard methods. A more elegant approach, however, exploits the fact that each of these integrals contains a closed subloop of massive gluons. This subloop 
admits the following dispersion representation 
\begin{equation}
\begin{aligned}
    \hspace{-3mm} \int \! \frac{\rmd^d k_1}{(2\pi)^d} \frac{1}{(k_1^2 - \mg^2)}
    \frac{1}{[(k_1+k_2)^2 - \mg^2]}
    = \frac{\rmi}{(4\pi)^{d/2}} \frac{\Gamma(1-\ep)}{\Gamma(2-2\ep)}
    \int \limits_{\mathclap{4\mg^2}}^\infty \ds \, \frac{s^{-\ep} (1 - 4\mg^2/s)^{\frac12 - \ep}}{s - k_2^2 - \rmi 0} \,,
    \label{eq_I6_dispersion_relation}
\end{aligned}
\end{equation}
which is very helpful 
for computing the three remaining integrals. 
Indeed, the denominator 
on the right-hand side of the above equation 
resembles the propagator of a particle with mass $s$, which means that the integration over $k_2$ in $I_{\rmB,\rmC,\rmD}$ can be easily performed. 

We start with the integral $I_\rmB$. 
Replacing the integral over $k_1$ in \eq\eqref{eq_I6_def} with
the dispersion representation in \eq\eqref{eq_I6_dispersion_relation}, we obtain 
\begin{equation}
    I_\rmB
    = \frac{-\rmi}{(4\pi)^{d/2}} \frac{\Gamma(1-\ep)}{\Gamma(2-2\ep)}
    \int \limits_{4\mg^2}^\infty \ds \, s^{-\ep} (1 - 4\mg^2/s)^{\frac12 - \ep} 
    \int \frac{\rmd^d k_2}{(2\pi)^d} \, \frac{1}{(k_2^2 - s)} \frac{1}{(2k_2\cdot p)} \,.
    \label{eq_I6_written_with_dispersion_relation}
\end{equation}
The integral over $k_2$ can be extracted from \eq\eqref{eq_F_of_s_def} after the replacement $\mg \to \sqrt{s}$.
We find 
\begin{equation}
    I_\rmB = \frac{1}{(4\pi)^{d}} \frac{\Gamma(1/2) \Gamma(1-\ep) \Gamma(-1/2+\ep)}{2\mt \Gamma(2-2\ep)} 
    \int \limits_{4\mg^2}^\infty \ds \, s^{\frac12-2\ep} (1 - 4\mg^2/s)^{\frac12 - \ep} \,.
\end{equation}
The integration over $s$ is straightforward. Indeed,  after changing the integration variable to $s \to 4\mg^2/x$, 
$I_{\rm B}$ can be easily evaluated  in terms of Gamma functions.
We obtain
\begin{equation}
    I_\rmB = \frac{\mg^{3-4\ep}}{\mt} \frac{2^{d-5}}{(4\pi)^{d-1}} \frac{\Gamma (-1/2 + \ep) \Gamma (-3/2 + 2 \ep)}{\Gamma (\ep )} \,.
\end{equation}

Turning now to the integral $I_\rmC$, we employ the dispersion relation in \eq\eqref{eq_I6_dispersion_relation} and find an expression analogous to \eq\eqref{eq_I6_written_with_dispersion_relation}, where now the integration over the loop momentum $k_2$ reads
\begin{equation}
    \int \frac{\rmd^d k_2}{(2\pi)^d} \, 
    \frac{1}{(k_2^2 - \mg^2)}
    \frac{1}{(k_2^2 - s)}
    \frac{1}{(2k_2\cdot p)} \,.
\end{equation}
The integration over $k_2$ becomes straightforward 
once we perform the partial fractioning 
\begin{equation}
    \frac{1}{(k_2^2 - \mg^2)} \frac{1}{(k_2^2 - s)}
    = \frac{1}{s-\mg^2} \bigg(\frac{1}{k_2^2 - s} - \frac{1}{k_2^2 - \mg^2}\bigg) \,,
\end{equation}
leading to
\begin{equation}
    I_\rmC = \frac{1}{(4\pi)^d} \frac{\Gamma(1/2) \Gamma(1-\ep) \Gamma(-1/2+\ep)}{2\mt\Gamma(2-2\ep)}
    \int \limits_{4\mg^2}^\infty \ds \, 
    \frac{s^{-\ep} (1 - 4\mg/s)^{\frac12 - \ep}}{s-\mg^2} \big(s^{\frac12 - \ep} - \mg^{1-2\ep} \big) \,.
    \label{eq_computation_of_F2_mid_step}
\end{equation} 
In the remaining integral over $s$, we again change the integration variable as $s \to 4\mg^2/x$, and write the integral as a difference of two hypergeometric functions. They  can be easily expanded in $\ep$ using the package HypExp \cite{Huber:2005yg,Huber:2007dx}. 
We find 
\begin{equation}
    \int \limits_{4\mg^2}^\infty \ds \, \frac{s^{-\ep} (1-4\mg^2/s)^{\frac12 - \ep}}{s-\mg^2} \big(s^{\frac12 - \ep} - \mg^{1-2\ep} \big)
    = - \mg^{1-4\ep} \left[\frac{1}{\ep} + \frac{2\pi}{\sqrt{3}} + \order{\ep} \right]\,.
\end{equation}
Combining the above pieces, we obtain 
\begin{equation}
    I_\rmC = \frac{\mg^{1-4\ep}}{\mt} \frac{\pi \Gamma^2(1+\ep)}{(4\pi)^d} 
    \left[\frac{1}{\ep} -2 \log{2} + \frac{2\pi}{\sqrt{3}} + 4 + \order{\ep^2} \right] \,.
\end{equation}

Finally, the integral $I_\rmD$ in \eq\eqref{eq_I8_def} can also be computed using the dispersion representation in  \eq\eqref{eq_I6_dispersion_relation}. 
The only difference is that, in this case, the one-loop 
bubble involves massless partons; therefore, the gluon mass $\mg$ in \eq\eqref{eq_I6_dispersion_relation} should be set to zero in both sides of the equation.
Following the steps described for $I_\rmC$, we obtain 
\begin{equation}
    I_\rmD = \frac{1}{(4\pi)^d} \frac{\Gamma(1/2) \Gamma(1-\ep) \Gamma(-1/2+\ep)}{2\mt\Gamma(2-2\ep)}
    \int \limits_{0}^\infty \ds \, 
    \frac{s^{-\ep} }{s-\mg^2} \big(s^{\frac12 - \ep} - \mg^{1-2\ep} \big) \,.
\label{eq_I8_dispersion_relation}
\end{equation} 
It is straightforward to compute the integral in the above equation as a series in $\ep$. 
We find\footnote{We note that this integral is known exactly. See, for instance, Ref.~\cite{gradshteyn2007}, Section 3.231, Eq.~(5).} 
\begin{equation}
    \int \limits_{0}^\infty \ds \, 
    \frac{s^{-\ep} }{s-\mg^2} \big(s^{\frac12 - \ep} - \mg^{1-2\ep} \big)
    =
    - \mg^{1-4\ep} 
    \left[\frac{1}{\ep} + \frac{5 \pi^2}{3} \ep 
    + \order{\ep} \right] \,.
\end{equation}
Replacing this expression in \eq\eqref{eq_I8_dispersion_relation}, we obtain the final result for $I_\rmD$,
\begin{equation}
    I_\rmD = \frac{\mg^{1-4\ep}}{\mt} \frac{\pi \Gamma^2(1+\ep)}{(4\pi)^d} 
    \left[\frac{1}{\ep} + 4 - 2\log{2} + \order{\ep} \right] \,.
\end{equation}


\subsection{Relation between the pole and the $\overline{\rm MS}$ top-quark masses}
\label{sec:relation-for-masses}

We are now in a position to discuss the relation between the pole mass and the $\MS$ mass of the top quark. Expanding the ratio of two 
$Z$ factors in the first relation in \eq\eqref{eq:the-relations} through $\OO(\as^2)$, we find 
\ali{
\label{eq:Zm-rel-expanded}
\!\! \frac{\mt^{{\rm \overline{MS}}}(\mu)}{\mt}\bigg|_{\OO(\as^2)} \!\!\!\!
=
\frac{Z_{\mt}^{\rm OS}}{Z_{\mt}^{\MS}(\mu)} \bigg|_{\OO(\as^2)} \!\!\!\!
=
\left(\delta Z_{\mt}^{\MS(1)}\right)^2 \! - \delta Z_{\mt}^{\MS(2)} - \delta Z_{\mt}^{\MS(1)} \delta Z_{\mt}^{\OS(1)} + \delta Z_{\mt}^{\OS(2)} \,. 
}
Recall that we are only interested in computing terms that are linear in the gluon mass. 
Since $Z_{\mt}^{\MS}$ is designed to remove only the UV divergences from the bare mass, it cannot contain any terms linear in $\mg$. This immediately implies that the first two terms on the right-hand side of \eq\eqref{eq:Zm-rel-expanded} do not contribute at $\order{\mg}$, so that 
\begin{equation}
    \label{eq:Zm-rel-expanded-var}
    \frac{\mt^{{\rm \overline{MS}}}(\mu)}{\mt}\bigg|_{\OO(\as^2,\mg)} \!\!\!\!
    = \frac{Z_{\mt}^{\rm OS}}{Z_{\mt}^{\MS}(\mu)} \bigg|_{\OO(\as^2,\mg)} \!\!\!\!
    = - \left( \delta Z_{\mt}^{\MS(1)} \delta Z_{\mt}^{\OS(1)}\right) \! \Big|_{\order{\mg}} \! + \delta Z_{\mt}^{\OS(2)}\big|_{\order{\mg}} \,.
\end{equation}
Since $\delta Z_{\mt}^{\MS(1)}$ itself contains no $\order{\mg}$ terms, the $\order{\mg}$ contribution in the first term on the right-hand side of \eq\eqref{eq:Zm-rel-expanded-var} must arise from the $\OO(\mg)$ term of $\delta Z_{\mt}^{\OS(1)}$. In any case, this term involves only one-loop counterterms and is therefore straightforward to compute. 
The second term, by contrast, is nontrivial; some technical details of its calculation have been already discussed in Sections \ref{sec:preliminaries} and \ref{sec:masters}. 
A few additional remarks are in order. 

First, we note that various contributions to $\delta Z_{\mt}^{\OS(2)}$
exhibit patterns that help 
to organize the calculation and clarify the structure of the result. 
In essence, they concern  cancellations of various contributions 
in \eq\eqref{eq:Zm-rel-expanded-var}, which happen because we are  only  interested in ${\cal O}(\mg)$ terms. Indeed, although the different contributions to \eq\eqref{eq:Zm-rel-expanded-var} have a cumbersome structure if the full dependence on $\mg$ is kept, our focus on ${\cal O}(\mg)$ terms leads to the appearance of simplifying patterns among them. 
These patterns are specific to leading-power soft contributions and, since the hard-hard region is irrelevant at  ${\cal O}(\mg)$, they are inherited by the final result at $\order{\mg}$.

One class of diagrams which exhibit such simplifications involves the one-loop vacuum polarization, which contributes not only to diagram (A) of Figure \ref{fig:bare}, but also to the counterterms and the diagrams that involve them, see  Figure~\ref{fig:CTs}. To better understand how this happens, we show the diagrams contributing to the vacuum polarization in Figure \ref{fig:VP-global}, separated into four classes: 
\begin{figure}[t!]
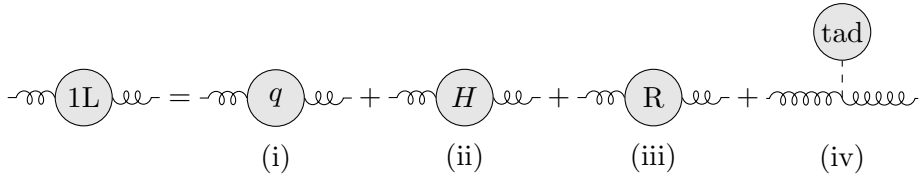

\centering
\vspace{-5mm}
\begin{align*}
 \FigTwoPointBlob{\rm 1L}{gluon}{gluon}
 = \FigTwoPointBlobWithLabelQ{gluon}{gluon}{(i)}
 + \FigTwoPointBlobWithLabel{H}{gluon}{gluon}{(ii)}
 + \FigTwoPointBlobWithLabel{\rm R}{gluon}{gluon}{(iii)}
 + \FigTwoPointBlobTadpoleWithLabel{(iv)}
\end{align*}
\vspace{-1.3em}
\caption{One-loop vacuum polarization.}
\label{fig:VP-global}
\end{figure}
(i) diagrams with a closed quark loop, (ii) 1-particle irreducible (1PI) diagrams involving the Higgs boson, (iii) the remaining 1PI diagrams, (iv) diagrams containing a one-loop Higgs tadpole. The specific Feynman diagrams contributing to each class are shown in Figure \ref{fig:VP-specific}. This separation into four classes is relevant because the contribution of a certain class to diagram (A) is intimately connected with the contribution of the same class to the counterterm diagrams. 
More specifically, we verified that the $\mg/\epsilon$ terms originating from the contribution of a certain class to diagram (A) cancel precisely against the $\mg/\epsilon$ terms coming from the contribution of the same class to $\delta Z^{(1)}_\rmg$ and $\delta Z^{(1)}_{\mg^2}$. We should add that there are cancellations even beyond the poles. In fact, for classes (ii), (iv) and the top-quark contribution to class (i), finite $\order{\mg}$ contributions to diagram (A) cancel  against similar
contributions 
to the counterterm diagrams.
\begin{figure}[t!]
 \centering
 \vspace{-5mm}
 \begin{gather*}
 \FigTwoPointBlobQ{gluon}{gluon} 
 = \FigTwoPointOneLoop{fermion}{fermion}{\psi_\rmt}{}
 + \FigTwoPointOneLoop{fermion}{fermion}{\psi_k}{}
 \end{gather*} 
 \vspace{-3em}
 \begin{gather*}
 \FigTwoPointBlob{H}{gluon}{gluon} 
 = \hspace{-0.5cm} \FigTwoPointSimpleTadpole{scalar}{H} \hspace{-0.5cm}
 + \FigTwoPointOneLoop{gluon}{scalar}{}{H}
 + \FigTwoPointOneLoop{scalar}{scalar}{H}{\varphi_a}
 \end{gather*}
 \vspace{-3em}
 \begin{gather*}
 \FigTwoPointBlob{\rm R}{gluon}{gluon} 
 = \hspace{-0.6cm} \FigTwoPointSimpleTadpole{gluon}{} \hspace{-0.6cm}
 + \hspace{-0.6cm} \FigTwoPointSimpleTadpole{scalar}{\varphi_a} \hspace{-0.6cm}
 + \FigTwoPointOneLoop{gluon}{gluon}{}{}
 + \FigTwoPointOneLoop{scalar}{scalar}{\varphi_a}{\varphi_b}
 + \FigTwoPointOneLoop{ghostArrow}{ghostArrow}{c_a}{c_b}
 \end{gather*}
 \vspace{-2em}
 \begin{gather*}
 \FigTwoPointBlobTadpole 
 = \FigTwoPointBalloonTadpole{gluon}{}
 + \FigTwoPointBalloonTadpole{scalar}{H}
 + \FigTwoPointBalloonTadpole{scalar}{\varphi_a}
 + \FigTwoPointBalloonTadpole{ghostArrow}{c_a}
 \end{gather*}
 \vspace{-1.3em}
 \caption{Details of the subsets of diagrams appearing in the vacuum polarization, cf.\ Figure \ref{fig:VP-global}.}
 \label{fig:VP-specific}
\end{figure}

Another useful organizing principle for the calculation is provided by the structure of color factors. Since different diagrams contribute with different Casimir operators, keeping track of these structures offers a helpful guide and provides nontrivial cross-checks. 
For example, only a few terms contribute with the color factor $\Cf^2$: diagrams (C) and (D) in Figure~\ref{fig:bare} (the former also contains a $\Ca \Cf$ contribution), diagram (d) in Figure~\ref{fig:CTs}, and the first term on the right-hand side of \eq\eqref{eq:Zm-rel-expanded-var}. 
We verified that the $\Cf^2 \mg/\epsilon$ terms from these contributions cancel among themselves. A similar pattern appears for the $\Ca \Cf$ terms that are not part of the vacuum polarization discussed in the previous paragraph.  
These include terms arising from diagram (C) and  diagram (B) 
of Figure~\ref{fig:bare}  as well 
as contributions from the counterterm $\delta Z_\rmg$.
We again verified the cancellation of the $\Ca \Cf \mg/\epsilon$ terms that originate from there. Finally, we note the peculiar fact that diagram (B), 
involving the three-gluon vertex, does not provide a double-soft contribution to the relation between the pole and $\MSbar$ masses, but it does contribute  to the relation between field renormalization constants.

A final remark concerns the one-loop counterterms. 
We checked that $\delta Z^{(1)}_\rmG$ and $\delta Z^{(1)}_{\psi_\rmt}$ do not contribute to the final result for the $\OSS$ mass counterterm. 
Also worth noting is the dependence of the one-loop counterterms on $\mg$. Interestingly, while we focus on the $\order{\mg}$ terms of the two-loop mass and field renormalization constants of the top quark, this requires the computation of $\delta Z_{\mg^2}^{(1)}$, which starts at $\order{\mg^{-2}}$, up to $\order{\mg}$ and $\delta Z_{\mt,\psi_\rmt}^{(1)}$ up to $\order{\mg}$. For $\delta Z_{\rmg}^{(1)}$ -- and therefore $\delta Z_\rmG^{(1)}$, which contributes to it -- we only need $\order{\mg^0}$ terms. 

Having discussed these aspects, we can finally present the relation between the pole mass and the $\MS$ mass of the top quark in the toy theory 
defined in Section~\ref{sec:model}.
We write it as
\be
\begin{split} 
\label{eq:final-mass-relation}
 \mt^{{\rm \overline{MS}}}(\mu) & = \mt f_1\big(\as(\mu),\mu, \mt, \mh\big) + 
 \as(\mu) \mg \frac{\Cf}{2} \bigg\{
 1 + \frac{\as(\mu)}{2\pi}
 \bigg[\Cf \bigg(1 + 3 \log\!\frac{\mt}{\mu}\bigg)
 \\
 & + \Ca \bigg(\frac{21 \sqrt{3} \pi}{32} - \frac{19}{48} - \frac{29}{8} \log\!\frac{\mg}{\mu}\bigg)
 - \frac{4}{9} \nf\TR \bigg(1- 3 \log\frac{\mg}{\mu} \bigg)
 \bigg] \bigg\} 
 +{\cal O}(\mg^2)\,.
\end{split} 
\ee
The function $f_1$ is independent of the gluon mass and therefore is of no interest to us.

The second, ${\cal O}(\mg)$, term on the right-hand side 
of \eq\eqref{eq:final-mass-relation}
is the main result of this paper.  
It  explicitly confirms the expectation that the relation between the $\MSbar$ and pole masses of a heavy quark has a linear dependence on the gluon mass $\mg$. 
The appearance of such a term supports the assertion that the toy 
model of Section~\ref{sec:model} provides a consistent framework to probe the IR sensitivity of observables through the parameter $\mg$. 
The $\OO(\as^2 \mg)$ term has a rich structure, depending on the combinations of various color factors, including the $\Ca \Cf$ term that reveals the contribution of non-Abelian dynamics. 

The dependence of the ${\cal O}(\mg)$ term in \eq\eqref{eq:final-mass-relation} 
 on the renormalization scale 
 $\mu $
  is also quite instructive. We recall 
that the $\mu$-dependence of the $\MS$ mass is dictated by the
renormalization group equation. Accordingly, the $\Cf^2 \log \mt/\mu$ term is needed in Eq.~(\ref{eq:final-mass-relation}) to ensure the correct renormalization group running of $\mt^{{\rm \overline{MS}}}(\mu)$. At the same time, the logarithms of $\mg/\mu$ can be absorbed into the running 
of the coupling constant. 
All of this suggests that \eq\eqref{eq:final-mass-relation} can be rewritten in a more elegant way by expressing the pole mass in terms of the $\MS$ mass. To the accuracy we work with, we only need the one-loop contribution to function $f_1$; 
it reads 
\be
f_1\big(\as(\mu),\mu,\mt,\mh\big) = 1 + \frac{\as(\mu)}{\pi} \Cf
\left ( \frac{3}{2} \log \frac{\mt}{\mu} -1 
\right ) + \OO(\as^2) \,.
\ee
Defining $\bar f_1$ as $1/f_1$ expanded in powers of $\as$, we readily find 
\begin{equation}
\begin{aligned}
 \mt = &\; \mt^{{\rm \overline{MS}}}(\mu)
 \bar f_1 \left (\as,\mu,\mt^{{\rm \overline{MS}}}(\mu),\mh 
 \right ) \\ 
 & - \as(\mg) \, \mg
 \frac{\Cf}{2} 
 \bigg\{1+
 \frac{ \as(\mg)}{2 \pi} 
 \bigg[ 3 \Cf + \Ca 
 \bigg(\frac{21 \sqrt{3} \pi}{32} - \dfrac{19}{48} \bigg) - \frac{4}{9} \nf \TR \bigg]
 \bigg\} \,.
\label{eq:final-mass-relation-2}
\end{aligned}
\end{equation}
This result clearly shows that the 
${\cal O}(\mg)$ correction to top quark pole mass
is entirely determined by energy scales comparable to the mass of the gluon.\footnote{
We note that the gluon mass $m_g$ should be chosen sufficiently large to make  perturbative computations sensible.}

As a final comment, we note that the pole mass, being a physical parameter, must be gauge independent. To verify this, we calculated the expression for the pole mass in a general $R_{\xi}$ gauge. We checked that the dependence on the $\xi$ parameter cancels, which provides a powerful check of our result.%
\fn{Interestingly, working in a general $R_{\xi}$ gauge leads to significantly more complicated soft master integrals, since the Goldstone bosons acquire the mass $\sqrt{\xi} \mg$. 
However, all nontrivial master integrals cancel in the final expression for $\delta Z_{\mt}^{\OS(2)}\big|_{\order{\mg}}$ and therefore do not need to be evaluated explicitly.}

\subsection{Relation between the top-quark field counterterms
in the on-shell subtraction and $\overline{\rm MS}$ schemes
}
\label{sec:relation-for-field}

We now briefly discuss the relation between field counterterms in the $\MS$ and $\OSS$ schemes, through two loops. 
This discussion will be concise, since our main focus in this work is the relation between the pole and $\MS$ masses. As explained in \Sec\ref{sec:introduction}, the 
field counterterm is needed for studying the IR sensitivity of 
observables in collider processes with heavy quarks.

Two aspects are worth mentioning. First, as discussed in \Sec\ref{sec:introduction}, this relation contains not only a linear, but also a logarithmic dependence on $\mg$.
Second, the $\OSS$  field counterterm is known to be gauge dependent. We computed the relation between the field counterterms in a general $R_{\xi}$ gauge and explicitly verified this property. For simplicity, we present the result in Feynman gauge. 
It reads
\begin{equation}
\begin{aligned}
 \sqrt{Z_{\psit}^{\OS}} = &\; \sqrt{Z_{\psit}^{\MS}}(\mu_0)\bigg\{
 f'\left(\as,\mu,\mt,\mh,\log \mg\right) + \frac{3}{8} \Cf\,\as(\mg) \frac{\mg}{\mt} \times \\
 & \times
 \bigg[
 1+\frac{\as(\mg)}{2\pi}
 \left(
 \Cf - \Ca\left(\frac{17}{16}-\frac{31\pi}{32\sqrt{3}} - \log\frac{\mt}{\mg}\right) - \frac{4}{9} \nf \TR
 \right)
 \bigg]
 \bigg\} \,,
\end{aligned}
\label{eq:psi-OS}
\end{equation}
where
\begin{equation}
 \sqrt{Z_{\psit}^{\MS}}(\mu_0) = \sqrt{Z_{\psit}^{\MS}} (\mu) \bigg[1 - \frac{\as(\mu) \Cf}{4 \pi} \log \frac{\mu}{\mu_0} \bigg] \,, 
 \qquad 
 \mu_0=\frac{\mt^3}{\mg^2} \,.
\end{equation}
We note that the scale $\mu_0$, 
which involves a peculiar combination of the top quark mass and the gluon mass, reflects the fact that 
the $\OSS$ renormalization constant $Z_{\psit}^{\OS}$ exhibits logarithmic sensitivity to \emph{both} 
UV and IR physics. 
Similarly to the function $f$, the function $f'$ has no linear dependence on $\mg$ and is therefore of no interest to us. One can observe that the $\order{\mg}$ term in \eq\eqref{eq:psi-OS} is explicitly $\mu$ independent. Notably, due to the $\log \mg$ term contained in $f'$, the $\OO(\as^2 \mg)$ term contains a logarithm of the ratio $\mt/\mg$ as well. 

We conclude this section noting that, for the investigation of the $\OO(\mg)$ contributions to heavy quark observables in collider processes, it is convenient to present the explicit expression for $Z^{\OS}_{\mt}$ and $\sqrt{Z_{\psit}^{\OS}}$, including poles and $\OO(\as \mg^0)$ contributions. We provide such expressions in Appendix~\ref{app:expressions}.
\section{Conclusions}
\label{sec:conclusions}

The steadily increasing precision of measurements in high-energy particle physics makes it essential to better understand the role of non-perturbative effects 
in processes with large momentum transfer. 
Since no reliable theoretical framework for this purpose exists,  
one can try to understand the magnitude of such effects by 
assessing the sensitivity of perturbative calculations to IR physics. 
From this perspective, the cancellation of logarithmic sensitivity to long-distance effects in IR-safe observables -- a direct consequence of the celebrated Bloch-Nordsieck and Kinoshita-Lee-Nauenberg theorems
-- ensures that perturbative predictions are valid with ${\cal O}(\Lambda_{\rm QCD}^0)$ accuracy, and that all non-perturbative corrections should be power-suppressed. An interesting question is whether a similar result, perhaps in a weaker form, can be also claimed for ${\cal O}(\Lambda_{\rm QCD})$ terms. 

A commonly used way to probe such sensitivity is provided by renormalon-based approaches. In this framework, the impact of long-distance physics is inferred from the behavior of perturbative corrections associated with the running of the strong coupling constant. In practical calculations, this sensitivity can be exposed by introducing a small gluon mass and studying how observables depend on it. Although this method has proven to be very useful, its scope is limited: introducing a gluon mass directly into QCD is only consistent in an Abelianized approximation and therefore does not allow one to study processes with external gluons.

Inspired by the renormalon 
approach, in this paper  we  promoted the gluon mass to a parameter of a consistent non-Abelian quantum field theory where 
controllable perturbative computations are possible beyond one loop. 
Specifically, we constructed a renormalizable toy model based on an SU(2) gauge theory whose symmetry is spontaneously broken through the Higgs mechanism. Within this setup, gluons acquire a small mass in a theoretically consistent manner, and this mass can be used as a probe of the IR sensitivity of perturbative observables. The resulting model provides a useful laboratory for investigating, in an unambiguous way, whether 
IR-safe observables studied at colliders exhibit linear  
IR sensitivity. 

As a first application, we analyzed the pole mass
of a heavy quark, which 
is a classic example of 
a quantity with linear IR sensitivity.
Focusing on the terms linear in the gluon mass, we computed 
the relation between the pole 
and the $\MSbar$ masses 
through the two-loop order, where genuinely non-Abelian effects appear for the first time. We discussed a number of subtleties that arise in the calculation, and derived the result for the linear contribution to the pole mass. 
This result is both gauge independent and renormalization-group invariant, providing a nontrivial check of the consistency of the framework. For completeness, we also derived the relation between the $\MS$ and on-shell subtraction field counterterms of the heavy quark.

The framework developed in this paper and the calculations presented here represent only the first steps toward a broader analysis of IR sensitivity of  collider observables. A natural next step is the study of heavy-quark pair production within this framework, which will provide a more direct connection to realistic processes relevant for precision measurements at hadron colliders.


\acknowledgments
This research was supported
by the Deutsche Forschungsgemeinschaft (DFG, German Research Foundation) under grant 396021762 - TRR 257. All Feynman diagrams and Feynman rules were drawn using \texttt{Tikz-feynman} \cite{Ellis:2016jkw}.

\appendix
\section{Feynman rules}
\label{App_Feynman_rules}

In this Appendix, we list the complete set of Feynman rules for the renormalized interactions of the toy model discussed in Section~\ref{sec:model}, for an arbitrary $R_{\xi}$ gauge (with the Feynman gauge corresponding to $\xi=1$). 
All momenta are taken to be incoming. At the end, we show a selected set of Feynman rules for the one-loop counterterms, relevant for the calculations discussed in the main text. 
The Feynman rules -- both for the renormalized interactions and the counterterms -- were generated in an automated way by FeynMaster \cite{Fontes:2019wqh,Fontes:2021iue,Fontes:2025svw}.
\vspace{-1.5mm}
\begin{align*}
    \intertext{\centering \underline{PROPAGATORS}} \\[-2.5em]
        \multicolumn{4}{c}{
        \FigPropagator{1}{G_\mu^a}{G_\nu^b}{gluon}
        $\hspace{-3mm} = - \dfrac{\rmi \delta_{ab}}{p^2 - \mg^2}\bigg[g_{\mu\nu} - (1-\xig)\dfrac{p_\mu p_\nu}{p^2 - \xig \mg^2}\bigg] $
        } \\
        &
        \FigPropagator{1}{\psi_{i}}{\psi_{j}}{fermion}
        \hspace{-2mm} = \dfrac{\rmi \delta_{ij} (\slashed{p} + m_\psi)}{p^2 - m_\psi^2} \,,
        && 
        \FigPropagator{1}{H}{H}{scalar}
        \hspace{-2mm} = \frac{\rmi}{p^2 - \mh^2}  \\ 
        &
        \FigPropagator{1}{\varphi_a}{\varphi_b}{scalar}
        \hspace{-2mm} = \frac{\rmi\delta_{ab}}{p^2 - \xig \mg^2} 
        && 
        \FigPropagator{1}{c_a}{c_b}{ghostArrow}
        \hspace{-2mm} = \frac{\rmi\delta_{ab}}{p^2 - \xig \mg^2} \\ 
    \intertext{\centering \underline{GLUON SELF INTERACTIONS}} \\[-3em]
        \multicolumn{4}{c}{
        \FigThreePointVerex{1}{G_\mu^a}{G_\nu^b}{G_\rho^c}{gluon}{gluon}{gluon}
        $\hspace{-3mm} = g \ep^{abc} \big[g_{\mu\nu}(p_b - p_a)^\rho
            + g_{\nu\rho}(p_c - p_b)^\mu
            + g_{\rho\mu}(p_a - p_c)^\nu\big] $
        } \\
        \multicolumn{4}{c}{ 
        \FigFourPointVerex{1}{G_\mu^a}{G_\nu^b}{G_\rho^c}{G_\sigma^d}{gluon}{gluon}{gluon}{gluon}
        $
        \begin{aligned}
            & \hspace{-4mm} =  \rmi g^2 
            \big[\ep^{abe}\ep^{cde} (g_{\mu\sigma} g_{\nu\rho} - g_{\mu\rho} g_{\nu\sigma})
            + \ep^{ace}\ep^{bde} (g_{\mu\sigma} g_{\nu\rho} - g_{\mu\nu} g_{\rho\sigma}) \\
            & + \ep^{ade}\ep^{bce} (g_{\mu\rho} g_{\nu\sigma} - g_{\mu\nu} g_{\rho\sigma})\big]
        \end{aligned}$
        } \\
    \intertext{\centering \underline{3-POINT INTERACTIONS}} \\[-2.5em]
        & \FigThreePointVerex{1}{\barpsi_i}{\psi_j}{G_\mu^a}{fermion}{fermion}{gluon}
        \hspace{-2mm} = - \rmi g \, \gamma^\mu \hatT^a_{ij}
        && 
        \FigThreePointVerex{1}{G_\mu^a}{G_\nu^b}{H}{gluon}{gluon}{scalar} 
        \hspace{-2mm} = \rmi g \, \mg \delta_{ab} \, g_{\mu\nu} \\
        & \FigThreePointVerex{1}{G_\mu^a}{\varphi_b}{H}{gluon}{scalar}{scalar}
        \hspace{-2mm} = \frac{g}{2} \delta_{ab} (p_\rmH - p_b)_\mu 
        && \FigThreePointVerex{1}{G_\mu^a}{\varphi_b}{\varphi_c}{gluon}{scalar}{scalar}
        \hspace{-2mm} = \frac{g}{2} \ep^{abc} (p_b - p_c)_\mu \\ 
        & \FigThreePointVerex{1}{H}{H}{H}{scalar}{scalar}{scalar}
        \hspace{-2mm} = - \frac{3\rmi g}{2} \frac{\mh^2}{\mg}  
        && \FigThreePointVerex{1}{\varphi_a}{\varphi_b}{H}{scalar}{scalar}{scalar}
        \hspace{-2mm} = - \frac{\rmi g}{2} \frac{\mh^2}{\mg} \delta_{ab} \\ 
        &\FigThreePointVerex{1}{\barc_a}{c_b}{G_\mu^c}{ghostArrow}{ghostArrow}{gluon}
        \hspace{-2mm} = - g \, \ep^{abc} (p_b + p_c)_\mu
        && \FigThreePointVerex{1}{\barc_a}{c_b}{H}{ghostArrow}{ghostArrow}{scalar}
        \hspace{-2mm} = - \frac{\rmi g \xig \mg}{2} \delta_{ab} \\
        & \FigThreePointVerex{1}{\barc_a}{c_b}{\varphi_c}{ghostArrow}{ghostArrow}{scalar}
        \hspace{-2mm} = - \frac{\rmi g \xig \mg}{2} \ep^{abc}
        && \\
    \intertext{\centering \underline{4-POINT INTERACTIONS}}
        & \FigFourPointVerex{1}{G_\mu^a}{G_\nu^b}{H}{H}{gluon}{gluon}{scalar}{scalar}
        \hspace{-2mm} = \frac{\rmi g^2}{2} \delta_{ab} \, g_{\mu\nu}
        && \FigFourPointVerex{1}{G_\mu^a}{G_\nu^b}{\varphi_c}{\varphi_d}{gluon}{gluon}{scalar}{scalar}
        \hspace{-2mm} = \frac{\rmi g^2}{2} \delta_{ab} \delta_{cd} \, g_{\mu\nu} \\ 
        & \FigFourPointVerex{1}{H}{H}{H}{H}{scalar}{scalar}{scalar}{scalar}
        \hspace{-2mm} = - \frac{3\rmi g^2}{4} \frac{\mh^2}{\mg^2}
        && \FigFourPointVerex{1}{H}{H}{\varphi_c}{\varphi_d}{scalar}{scalar}{scalar}{scalar}
        \hspace{-2mm} = - \frac{\rmi g^2}{4} \frac{\mh^2}{\mg^2} \delta_{cd} \\ 
        & \FigFourPointVerex{1}{\varphi_a}{\varphi_b}{\varphi_c}{\varphi_d}{scalar}{scalar}{scalar}{scalar}
        \hspace{-2mm} = - \frac{\rmi g^2}{4}\frac{\mh^2}{\mg^2} (\delta_{ab}\delta_{cd} + \delta_{ac}\delta_{bd} + \delta_{ad}\delta_{bc}) \hspace{-2cm}
        && 
    \intertext{\centering \underline{ONE-LOOP COUNTERTERMS}}
        \multicolumn{4}{c}{
        \FigPropagatorWithStart{1}{G_\mu^a}{G_\nu^b}{gluon}{black}
        $\hspace{-3mm} = \rmi \delta_{ab} \Big[\mg^2 \, g_{\mu\nu} \left( \delta Z^{(1)}_{\mg^2} 
        + \delta Z^{(1)}_\rmG \right) -  \delta Z^{(1)}_\rmG \left( p^2 g_{\mu\nu} -p_\mu p_\nu  \right) \Big]$
        } \\
        \multicolumn{4}{c}{
        \FigPropagatorWithStart{1}{\barpsi_{\rmt,i}}{\psi_{\rmt,j}}{fermion}{black}
        $\hspace{-3mm} = - \rmi \delta_{ij} \Big[\mt \delta Z^{(1)}_{\mt} - \delta Z^{(1)}_{\psit}(\slashed{p} - \mt) \Big]$
        } \\
        \multicolumn{4}{c}{
        \FigThreePointVerexWithStar{1}{\barpsi_{\rmt,i}}{\psi_{\rmt,j}}{G_\mu^a}{fermion}{fermion}{gluon}
        \hspace{-2mm} 
        $\hspace{-3mm} = -\rmi g \bigg[\delta Z^{(1)}_\rmg + \dfrac12 \delta Z^{(1)}_\rmG + \delta Z^{(1)}_{\psit} \bigg]\gamma_\mu \hatT^a_{ij}$  
        }
\end{align*}

\section{On-shell scheme conditions for 2-loop fermion mass and field counterterms}
\label{app:OS-conditions}

Ref.~\cite{Melnikov:2000zc} provides an efficient procedure to calculate counterterms in the OS scheme. In this appendix, with the goal of determining the nontrivial counterterms $\delta Z_{\mt}^{\OS(2)}$ and $\delta Z_{\psit}^{\OS(2)}$ contributing to the expanded version of \eq\eqref{eq:the-relations}, we rederive that procedure in the framework of renormalized perturbation theory. Accordingly, all calculations are performed with \textit{renormalized parameters}, while counterterm contributions are included through the corresponding Feynman rules. Since the counterterms in this appendix are all defined in the OS scheme, we simplify the notation by omitting the superscript OS. Finally, we denote renormalized loop functions with a hat and unrenormalized ones without it.

Before introducing the method of Ref.~\cite{Melnikov:2000zc}, we cover some preliminary aspects of renormalized perturbation theory. We write the renormalized 2-point function for the top quark as
\begin{equation}
    \label{eq:reno-Gamma}
    \hat{\Gamma}(\slashed{p})= \slashed{p}-\mt + \hat{\Sigma}(\slashed{p}) \,,
\end{equation}
where $\mt$ represents the renormalized top-quark mass, and $\hat{\Sigma}(\slashed{p})$ the renormalized higher-order contributions to the self-energy, which we parameterize as
\begin{equation}
    \label{eq:reno-Sigma}
    \hat{\Sigma}(\slashed{p})= \mt \, \hat{\Sigma}_1(p^2) + (\slashed{p}-\mt) \, \hat{\Sigma}_2(p^2) \,.
\end{equation}
The renormalized loop functions $\hat{\Sigma}_1$ and $\hat{\Sigma}_2$ are expanded order by order in perturbation theory as
\ali{
\label{eq:Sigma-expansion}
	\hat{\Sigma}_{1,2}(p^2) = \hat{\Sigma}_{1,2}^{(1)}(p^2) + \hat{\Sigma}_{1,2}^{(2)}(p^2) + ... \, .
}
The first term on the right-hand side represents the next-to-leading order term, the second one the next-to-next-to-leading order term, and the ellipses represent higher order terms, which are not relevant for our purposes.

To fix the nontrivial counterterms discussed above, we first need to relate the renormalized functions $\hat{\Sigma}_1^{(i)}(p^2)$ and $\hat{\Sigma}_2^{(i)}(p^2)$ to counterterms and unrenormalized functions. 
We start by writing the unrenormalized function $\Sigma(\slashed{p})$ in a form analogous to \eq\eqref{eq:reno-Sigma},
\ali{
    \label{eq:unrenormalized-Sigma}
    \Sigma(\slashed{p})= \mt \Sigma_1(p^2) + (\slashed{p}-\mt) \Sigma_2(p^2) \, .
}
We stress that the parameters involved in this equation are {renormalized}.
The unrenormalized loop functions $\Sigma_1(p^2)$ and $\Sigma_2(p^2)$ follow an expansion equivalent to \eq\eqref{eq:Sigma-expansion},
\ali{
\label{eq:Sigma-expansion-unrenormalized}
	\Sigma_{1,2}(p^2) = \Sigma_{1,2}^{(1)}(p^2) + \Sigma_{1,2}^{(2)}(p^2) + ... \, .
}
Now, we consider the terms in \eq\eqref{eq_Lagrangian} involving exclusively the top quark. After taking both the top-quark field and mass as bare, we relate them to their corresponding renormalized quantities and $Z$ factors through Eqs \eqref{eq:parameter-CTs} and \eqref{eq:field-CTs}. Finally, by expanding the $Z$ factors as described after \eq\eqref{eq:field-CTs}, and by using Eqs~\eqref{eq:reno-Gamma}--\eqref{eq:Sigma-expansion-unrenormalized}, we finally find
\begin{equation}
\begin{aligned}
    \hat{\Sigma}_{1}^{(1)}(p^2) &= \Sigma_1^{(1)}(p^2) - \delta Z_{\mt}^{(1)} \,, 
    &\qquad
    \hat{\Sigma}_{2}^{(1)}(p^2) &= \Sigma_2^{(1)}(p^2) + \delta Z_{\psit}^{(1)} \,, \\
    \hat{\Sigma}_{1}^{(2)}(p^2) &= \Sigma_1^{(2)}(p^2) - \delta Z_{\mt}^{(2)} - \delta Z_{\mt}^{(1)} \,\delta Z_{\psit}^{(1)} \,, 
    &\qquad
    \hat{\Sigma}_{2}^{(2)}(p^2) &= \Sigma_2^{(2)}(p^2) + \delta Z_{\psit}^{(2)} \,.
\label{eqs:unrenormalized-reno-rels}
\end{aligned}
\end{equation}

We can now fix the counterterms using OS conditions. We write the renormalized 2-point function in formal Taylor series around $p^2=\mt^2$, 
\begin{equation}
\begin{aligned}
    \hat{\Gamma}(\slashed{p})
    &\approx \mt\,\hat{\Sigma}_1(\mt^2)
    + \mt \, (p^2-\mt^2)  \, \frac{\partial \hat{\Sigma}_1(p^2)}{\partial p^2}\Big|_{p^2=\mt^2} + (\slashed{p}-\mt)\, \Big[1 + \hat{\Sigma}_2(\mt^2) \Big] \\
    &\approx \mt\,\hat{\Sigma}_1(\mt^2) + (\slashed{p}-\mt)\bigg[1+2\mt^2\frac{\partial \hat{\Sigma}_1(p^2)}{\partial p^2}\Big|_{p^2=\mt^2} + \hat{\Sigma}_2(\mt^2)
    \bigg] \,.
\end{aligned}
\end{equation}
From this expression, the well-known OS conditions relative to the pole and residue follow in a trivial way; they are, respectively,
\ali{
\hat{\Sigma}_1(\mt^2)=0 \,, 
\qquad
2\mt^2\frac{\partial \hat{\Sigma}_1(p^2)}{\partial p^2}\Big|_{p^2=\mt^2} + \hat{\Sigma}_2(\mt^2) = 0 \,,
}
which, combined with Eqs \eqref{eq:Sigma-expansion} and \eqref{eqs:unrenormalized-reno-rels}, respectively fix the mass and field counterterms to
\begin{align}
    & \delta Z_m^{(1)} = \Sigma_1^{(1)}(\mt^2) \,,
    \qquad 
    \delta Z_m^{(2)} = \Sigma_1^{(2)}(\mt^2) - \delta Z^{(1)}_{\mt} \delta Z^{(1)}_{\psit} \,, \label{eq_appendix_deltaZm}\\
    &
    \begin{aligned}
        \delta Z_{\psit}^{(i)} & = - \bigg[ 2\mt^2\frac{\partial \Sigma_1^{(i)} (p^2)}{\partial p^2}\Big|_{p^2=\mt^2} + \Sigma_2^{(i)}(\mt^2) \bigg] \,, 
        \quad i = 1,2 \,.
        \label{eq_appendix_deltaZpsit}
    \end{aligned}
\end{align}
Note that the functions $\Sigma_{1,2}^{(2)}$, which are unrenormalized two-loop functions, receive contributions not only from genuine two-loop diagrams (as in Figure~\ref{fig:bare}), but also from one-loop diagrams with counterterm insertions (as in Figure~\ref{fig:CTs}). 

The quantities $\delta Z_m^{(2)}$ and $\delta Z_{\psit}^{(2)}$ in Eqs \eqref{eq_appendix_deltaZm} and \eqref{eq_appendix_deltaZpsit} are the nontrivial counterterms to be determined.
Their calculation is simplified in the method of Ref.~\cite{Melnikov:2000zc} by
introducing a 4-vector $Q_{\mu}$ such that $Q^2=\mt^2$ and $p_{\mu} = Q_{\mu}(1+t)$, where $t$ is a small parameter. Then,
\ali{
\Sigma(\slashed{p})
= \mt \, \Sigma_1(p^2)
+ (\slashed{Q}- \mt)\,\Sigma_2(p^2)
+ t\,\slashed{Q}\,\Sigma_2(p^2) \,.
}
We now introduce the quantity $T_1$, defined as
\ali{
T_1
= \mathrm{Tr}\!\left[
\frac{\slashed{Q}+\mt}{4 \mt^2}\,\Sigma(\slashed{p})
\right]
= \Sigma_1(p^2) + t\,\Sigma_2(p^2) \,.
}
Expanding in $t$ up to corrections of $\order{t^2}$, we obtain
\ali{
T_1
= \Sigma_1(\mt^2)
+ t\left[
2 \mt^2\,\frac{\partial \Sigma_1(p^2)}{\partial p^2}\Big|_{p^2=\mt^2}
+ \Sigma_2(\mt^2)
\right]
+ \mathcal{O}(t^2) \,.
}
The quantity $T_1$ thus expanded is particularly convenient: focusing on the two-loop case, $\order{t^0}$ term corresponds to the first term of $\delta Z_m^{(2)}$, while the $\order{t}$ term yields $-\delta Z_{\psit}^{(2)}$.

\section{Expressions for the two-loop counterterms}
\label{app:expressions}

For practical applications, it is convenient to  write the expressions for $Z^{\OS}_{\mt}$ and $\sqrt{Z_{\psit}^{\OS}}$
with all their $1/\ep$ poles explicitly. 
Since those applications concern $\OO(\mg)$ contributions in two-loop calculations, the $\OO(\as)$ term of such expressions must be kept through $\OO(\epsilon)$ and $\OO(\mg)$, whereas at $\OO(\as^2)$ it suffices to retain the term linear in $\mg$ through $\OO(\epsilon^0)$. Hence, we write
\ali{
Z_{\mt}^{\OS} &= 1 + \frac{\Cf \as(\mu)}{2\pi} \bigg[c_{m}^{(1,0)} \! + \frac{\mg}{\mt} c_{m}^{(1,1)} + \as(\mu) \bigg( c_{m}^{(2,0)} \! + \frac{\mg}{\mt} c_{m}^{(2,1)}\bigg) + \order{\as^3,\mg^2} \bigg] \,, \\
\sqrt{Z_{\psit}^{\OS}} &= 1 + \frac{\Cf \as(\mu)}{2\pi} \bigg[c_{\psi}^{(1,0)} \! + \frac{\mg}{\mt} c_{\psi}^{(1,1)} + \as(\mu) \bigg( c_{\psi}^{(2,0)} \! + \frac{\mg}{\mt} c_{\psi}^{(2,1)}\bigg) + \order{\as^3,\mg^2} \bigg] \,,
}
where $c_X^{(i,j)}$ represents the coefficient of the term of $\OO(\as^i \mg^j)$.
They read, up to corrections of $\order{\ep^2}$,
\begin{align}
    \hspace{-2.5mm} 
    c_{m}^{(1,0)} =& - \frac{3}{2\epsilon} + 3\ln\frac{\mt}{\mu} - 2
    + \epsilon\left(- 3\ln^2\frac{\mt}{\mu}
    + 4\ln\frac{\mt}{\mu} - \frac{\pi^2}{8} -4
    \right) \,, \\
    c_{m}^{(1,1)} =&\; \pi - 2 \pi \epsilon \left( \ln \frac{2 \mg}{\mu} -1 \right) \,, \\
    c_{m}^{(2,1)} =&\; \frac{\Cf}{2}\left(
    -\frac{3}{2\epsilon}+3\ln\frac{2\mg}{\mu} + 3\ln\frac{\mt}{\mu} -2
    \right) + \frac{2}{9} \nf \TR \left(3\ln\frac{\mg}{\mu}-1 \right) \\
    & + \frac{\Ca}{192}\left(
    -38 + 63\sqrt{3}\,\pi - 348\ln\frac{\mg}{\mu}
    \right) \,, \notag \\
    c_{\psi}^{(1,0)} =&\; -\frac{1}{4 \epsilon} -1 + \frac{1}{2}\log\frac{\mt^3}{\mg^2 \mu} + \epsilon \bigg[ - \frac{3}{2}\log^2\frac{\mt}{\mu} + 2 \log \frac{\mt}{\mu} + \log^2\frac{\mg}{\mu} -2 - \frac{\pi^2}{48} \bigg] \,, \\
    c_{\psi}^{(1,1)} =&\; \frac{3\pi}{4} + \pi \epsilon\left[1 - \frac{3}{2}\log\frac{2\mg}{\mu}\right] \,, \\
    c_{\psi}^{(2,1)} =&\; \Cf\left(-\frac{3}{32 \epsilon} + \frac{1}{4} + \frac{3}{16}\log\frac{2\mt^3}{\mg \mu^2} \right) + \TR \nf\left(-\frac{1}{6} + \frac{1}{2}\log\frac{\mg}{\mu}\right) \\
    & + \Ca \bigg( -\frac{51}{128} + \frac{31\sqrt{3}\,\pi}{256} - \frac{111}{64}\log\frac{\mg}{\mu} + \frac{3}{8}\log\frac{\mt}{\mu} \bigg) \,. \notag
\end{align}
As usual, the coefficients $c_X^{(2,0)}$ are of no interest to us, as they does not lead to a linear dependence on $\mg$ at the two-loop level.


\bibliographystyle{JHEP}
\bibliography{biblio.bib}

@article{Sinkovics:1998mi,
    author = "Sinkovics, A. and Akhoury, R. and Zakharov, Valentin I.",
    title = "{Cancellation of $1 / m_Q$ corrections to the inclusive decay width of a heavy quark}",
    eprint = "hep-ph/9804401",
    archivePrefix = "arXiv",
    reportNumber = "UM-TH-98-08",
    doi = "10.1103/PhysRevD.58.114025",
    journal = "Phys. Rev. D",
    volume = "58",
    pages = "114025",
    year = "1998"
}

@article{Shtabovenko:2023idz,
    author = "Shtabovenko, Vladyslav and Mertig, Rolf and Orellana, Frederik",
    title = "{FeynCalc 10: Do multiloop integrals dream of computer codes?}",
    eprint = "2312.14089",
    archivePrefix = "arXiv",
    primaryClass = "hep-ph",
    reportNumber = "P3H-23-089, TTP23-056, SI-HEP-2023-27",
    doi = "10.1016/j.cpc.2024.109357",
    journal = "Comput. Phys. Commun.",
    volume = "306",
    pages = "109357",
    year = "2025"
}

@article{Shtabovenko:2025lxq,
    author = "Shtabovenko, Vladyslav",
    title = "{FeynCalc 10.2 and FeynHelpers 2: Multiloop calculations streamlined}",
    eprint = "2512.19858",
    archivePrefix = "arXiv",
    primaryClass = "hep-ph",
    reportNumber = "P3H-25-112, SI-HEP-2025-31",
    month = "12",
    year = "2025"
}

@article{Fontes:2019wqh,
    author = "Fontes, Duarte and Rom{\~a}o, Jorge C.",
    title = "{FeynMaster: a plethora of Feynman tools}",
    eprint = "1909.05876",
    archivePrefix = "arXiv",
    primaryClass = "hep-ph",
    doi = "10.1016/j.cpc.2020.107311",
    journal = "Comput. Phys. Commun.",
    volume = "256",
    pages = "107311",
    year = "2020"
}

@inproceedings{Fontes:2025svw,
    author = "Fontes, Duarte and Rom{\~a}o, Jorge C.",
    title = "{FeynMaster Manual}",
    eprint = "2504.01865",
    archivePrefix = "arXiv",
    primaryClass = "hep-ph",
    year = "2025"
}

@article{Christensen:2008py,
    author = "Christensen, Neil D. and Duhr, Claude",
    title = "{FeynRules - Feynman rules made easy}",
    eprint = "0806.4194",
    archivePrefix = "arXiv",
    primaryClass = "hep-ph",
    reportNumber = "MSUHEP-080616, CP3-08-20",
    doi = "10.1016/j.cpc.2009.02.018",
    journal = "Comput. Phys. Commun.",
    volume = "180",
    pages = "1614--1641",
    year = "2009"
}

@article{Alloul:2013bka,
    author = "Alloul, Adam and Christensen, Neil D. and Degrande, C{\'e}line and Duhr, Claude and Fuks, Benjamin",
    title = "{FeynRules  2.0 - A complete toolbox for tree-level phenomenology}",
    eprint = "1310.1921",
    archivePrefix = "arXiv",
    primaryClass = "hep-ph",
    reportNumber = "CERN-PH-TH-2013-239, MCNET-13-14, IPPP-13-71, DCPT-13-142, PITT-PACC-1308",
    doi = "10.1016/j.cpc.2014.04.012",
    journal = "Comput. Phys. Commun.",
    volume = "185",
    pages = "2250--2300",
    year = "2014"
}

@article{Fontes:2021iue,
    author = "Fontes, Duarte and Rom{\~a}o, Jorge C.",
    title = "{Renormalization of the C2HDM with FeynMaster 2}",
    eprint = "2103.06281",
    archivePrefix = "arXiv",
    primaryClass = "hep-ph",
    doi = "10.1007/JHEP06(2021)016",
    journal = "JHEP",
    volume = "06",
    pages = "016",
    year = "2021",
    note = "[Erratum: JHEP 12, 005 (2021)]"
}

@article{Nogueira:1991ex,
    author = "Nogueira, Paulo",
    title = "{Automatic Feynman Graph Generation}",
    reportNumber = "IFM-7-91",
    doi = "10.1006/jcph.1993.1074",
    journal = "J. Comput. Phys.",
    volume = "105",
    pages = "279--289",
    year = "1993"
}

@article{Mertig:1990an,
    author = "Mertig, R. and Bohm, M. and Denner, Ansgar",
    title = "{FEYN CALC: Computer algebraic calculation of Feynman amplitudes}",
    reportNumber = "PRINT-90-0639 (WURZBURG)",
    doi = "10.1016/0010-4655(91)90130-D",
    journal = "Comput. Phys. Commun.",
    volume = "64",
    pages = "345--359",
    year = "1991"
}

@article{Shtabovenko:2016sxi,
    author = "Shtabovenko, Vladyslav and Mertig, Rolf and Orellana, Frederik",
    title = "{New Developments in FeynCalc 9.0}",
    eprint = "1601.01167",
    archivePrefix = "arXiv",
    primaryClass = "hep-ph",
    reportNumber = "TUM-EFT-71-15",
    doi = "10.1016/j.cpc.2016.06.008",
    journal = "Comput. Phys. Commun.",
    volume = "207",
    pages = "432--444",
    year = "2016"
}

@article{Shtabovenko:2020gxv,
    author = "Shtabovenko, Vladyslav and Mertig, Rolf and Orellana, Frederik",
    title = "{FeynCalc 9.3: New features and improvements}",
    eprint = "2001.04407",
    archivePrefix = "arXiv",
    primaryClass = "hep-ph",
    reportNumber = "P3H-20-002, TTP19-020, TUM-EFT 130/19",
    doi = "10.1016/j.cpc.2020.107478",
    journal = "Comput. Phys. Commun.",
    volume = "256",
    pages = "107478",
    year = "2020"
}

@phdthesis{deSousaMachadoFontes:2021zia,
    author = "Fontes, Duarte",
    title = "{Multi-Higgs Models: model building, phenomenology and renormalization}",
    eprint = "2109.08394",
    archivePrefix = "arXiv",
    primaryClass = "hep-ph",
    school = "U. Lisbon (main)",
    year = "2021"
}

@article{Beneke:1997zp,
    author = "Beneke, M. and Smirnov, Vladimir A.",
    title = "{Asymptotic expansion of Feynman integrals near threshold}",
    eprint = "hep-ph/9711391",
    archivePrefix = "arXiv",
    reportNumber = "CERN-TH-97-315",
    doi = "10.1016/S0550-3213(98)00138-2",
    journal = "Nucl. Phys. B",
    volume = "522",
    pages = "321--344",
    year = "1998"
}

@article{Beneke:1999zr,
    author = "Beneke, M.",
    editor = "Sachrajda, Christopher",
    title = "{Perturbative heavy quark - anti-quark systems}",
    eprint = "hep-ph/9911490",
    archivePrefix = "arXiv",
    reportNumber = "CERN-TH-99-355",
    doi = "10.22323/1.003.0009",
    journal = "PoS",
    volume = "hf8",
    pages = "009",
    year = "1999"
}

@article{Jantzen:2011nz,
    author = "Jantzen, Bernd",
    title = "{Foundation and generalization of the expansion by regions}",
    eprint = "1111.2589",
    archivePrefix = "arXiv",
    primaryClass = "hep-ph",
    reportNumber = "TTK-11-53, SFB-CPP-11-61",
    doi = "10.1007/JHEP12(2011)076",
    journal = "JHEP",
    volume = "12",
    pages = "076",
    year = "2011"
}

@article{Patel:2015tea,
    author = "Patel, Hiren H.",
    title = "{Package-X: A Mathematica package for the analytic calculation of one-loop integrals}",
    eprint = "1503.01469",
    archivePrefix = "arXiv",
    primaryClass = "hep-ph",
    doi = "10.1016/j.cpc.2015.08.017",
    journal = "Comput. Phys. Commun.",
    volume = "197",
    pages = "276--290",
    year = "2015"
}

@article{Patel:2016fam,
    author = "Patel, Hiren H.",
    title = "{Package-X 2.0: A Mathematica package for the analytic calculation of one-loop integrals}",
    eprint = "1612.00009",
    archivePrefix = "arXiv",
    primaryClass = "hep-ph",
    doi = "10.1016/j.cpc.2017.04.015",
    journal = "Comput. Phys. Commun.",
    volume = "218",
    pages = "66--70",
    year = "2017"
}

@article{Huber:2005yg,
    author = "Huber, Tobias and Ma{\^\i}tre, Daniel",
    title = "{HypExp, a Mathematica package for expanding hypergeometric functions around integer-valued parameters}",
    eprint = "hep-ph/0507094",
    archivePrefix = "arXiv",
    reportNumber = "ZU-TH-13-05",
    doi = "10.1016/j.cpc.2006.01.007",
    journal = "Comput. Phys. Commun.",
    volume = "175",
    pages = "122--144",
    year = "2006"
}

@article{Huber:2007dx,
    author = "Huber, Tobias and Ma{\^\i}tre, Daniel",
    title = "{HypExp 2, Expanding hypergeometric functions about half-integer parameters}",
    eprint = "0708.2443",
    archivePrefix = "arXiv",
    primaryClass = "hep-ph",
    reportNumber = "SLAC-PUB-12748, PITHA-07-06",
    doi = "10.1016/j.cpc.2007.12.008",
    journal = "Comput. Phys. Commun.",
    volume = "178",
    pages = "755--776",
    year = "2008"
}

@article{Melnikov:2000zc,
    author = "Melnikov, Kirill and van Ritbergen, Timo",
    title = "{The Three loop on-shell renormalization of QCD and QED}",
    eprint = "hep-ph/0005131",
    archivePrefix = "arXiv",
    reportNumber = "SLAC-PUB-8450, TTP-00-08",
    doi = "10.1016/S0550-3213(00)00526-5",
    journal = "Nucl. Phys. B",
    volume = "591",
    pages = "515--546",
    year = "2000"
}

@book{gradshteyn2007,
  author = {Gradshteyn, I. S. and Ryzhik, I. M.},
  edition = {Seventh},
  isbn = {978-0-12-373637-6; 0-12-373637-4},
  mrclass = {00A22 (33-00 65-00 65A05)},
  mrnumber = {2360010 (2008g:00005)},
  note = {Translated from the Russian, Translation edited and with a preface by Alan Jeffrey and Daniel Zwillinger, With one CD-ROM (Windows, Macintosh and UNIX)},
  pages = {xlviii+1171},
  publisher = {Elsevier/Academic Press, Amsterdam},
  title = {Table of integrals, series, and products},
  year = 2007
}

@article{Beneke:1994sw,
    author = "Beneke, M. and Braun, Vladimir M.",
    title = "{Heavy quark effective theory beyond perturbation theory: Renormalons, the pole mass and the residual mass term}",
    eprint = "hep-ph/9402364",
    archivePrefix = "arXiv",
    reportNumber = "MPI-PHT-94-9, UM-TH-94-4",
    doi = "10.1016/0550-3213(94)90314-X",
    journal = "Nucl. Phys. B",
    volume = "426",
    pages = "301--343",
    year = "1994"
}

@article{Manohar:1994kq,
    author = "Manohar, Aneesh V. and Wise, Mark B.",
    title = "{Power suppressed corrections to hadronic event shapes}",
    eprint = "hep-ph/9406392",
    archivePrefix = "arXiv",
    reportNumber = "UCSD-PTH-94-11, CALT-68-1937",
    doi = "10.1016/0370-2693(94)01504-6",
    journal = "Phys. Lett. B",
    volume = "344",
    pages = "407--412",
    year = "1995"
}

@article{Bigi:1994em,
    author = "Bigi, Ikaros I. Y. and Shifman, Mikhail A. and Uraltsev, N. G. and Vainshtein, A. I.",
    title = "{The Pole mass of the heavy quark. Perturbation theory and beyond}",
    eprint = "hep-ph/9402360",
    archivePrefix = "arXiv",
    reportNumber = "TPI-MINN-94-4-T, UMN-TH-1239-94, CERN-TH-7171-94, UND-HEP-94-BIG03",
    doi = "10.1103/PhysRevD.50.2234",
    journal = "Phys. Rev. D",
    volume = "50",
    pages = "2234--2246",
    year = "1994"
}

@article{Caola:2021kzt,
    author = "Caola, Fabrizio and Ferrario Ravasio, Silvia and Limatola, Giovanni and Melnikov, Kirill and Nason, Paolo",
    title = "{On linear power corrections in certain collider observables}",
    eprint = "2108.08897",
    archivePrefix = "arXiv",
    primaryClass = "hep-ph",
    reportNumber = "OUTP-21-21P, TTP21-026, P3H-21-056",
    doi = "10.1007/JHEP01(2022)093",
    journal = "JHEP",
    volume = "01",
    pages = "093",
    year = "2022"
}

@article{Akhoury:1996ks,
    author = "Akhoury, R. and Stodolsky, Leo and Zakharov, Valentin I.",
    title = "{Power corrections and KLN cancellations}",
    eprint = "hep-ph/9609368",
    archivePrefix = "arXiv",
    reportNumber = "MPI-PH-96-88, UM-TH-96-14",
    doi = "10.1016/S0550-3213(97)00766-9",
    journal = "Nucl. Phys. B",
    volume = "516",
    pages = "317--332",
    year = "1998"
}

@article{Hoang:2008yj,
    author = "Hoang, Andre H. and Jain, Ambar and Scimemi, Ignazio and Stewart, Iain W.",
    title = "{Infrared Renormalization Group Flow for Heavy Quark Masses}",
    eprint = "0803.4214",
    archivePrefix = "arXiv",
    primaryClass = "hep-ph",
    reportNumber = "MIT-CTP-3940, MPP-2008-24",
    doi = "10.1103/PhysRevLett.101.151602",
    journal = "Phys. Rev. Lett.",
    volume = "101",
    pages = "151602",
    year = "2008"
}

@article{Beneke:1998rk,
    author = "Beneke, M.",
    title = "{A Quark mass definition adequate for threshold problems}",
    eprint = "hep-ph/9804241",
    archivePrefix = "arXiv",
    reportNumber = "CERN-TH-98-120",
    doi = "10.1016/S0370-2693(98)00741-2",
    journal = "Phys. Lett. B",
    volume = "434",
    pages = "115--125",
    year = "1998"
}

@article{Smirnov:2025prc,
    author = "Smirnov, Alexander V. and Zeng, Mao",
    title = "{FIRE 7: Automatic Reduction with Modular Approach}",
    eprint = "2510.07150",
    archivePrefix = "arXiv",
    primaryClass = "hep-ph",
    month = "10",
    year = "2025"
}

@article{Lee:2012cn,
    author = "Lee, R. N.",
    title = "{Presenting LiteRed: a tool for the Loop InTEgrals REDuction}",
    eprint = "1212.2685",
    archivePrefix = "arXiv",
    primaryClass = "hep-ph",
    month = "12",
    year = "2012"
}

@article{Korchemsky:1994is,
    author = "Korchemsky, Gregory P. and Sterman, George F.",
    title = "{Nonperturbative corrections in resummed cross-sections}",
    eprint = "hep-ph/9411211",
    archivePrefix = "arXiv",
    reportNumber = "ITP-SB-94-50, ITP--SB--94--50, NOVEMBER-1994",
    doi = "10.1016/0550-3213(94)00006-Z",
    journal = "Nucl. Phys. B",
    volume = "437",
    pages = "415--432",
    year = "1995"
}

@article{Webber:1994cp,
    author = "Webber, B. R.",
    title = "{Estimation of power corrections to hadronic event shapes}",
    eprint = "hep-ph/9408222",
    archivePrefix = "arXiv",
    reportNumber = "CAVENDISH-HEP-94-7",
    doi = "10.1016/0370-2693(94)91147-9",
    journal = "Phys. Lett. B",
    volume = "339",
    pages = "148--150",
    year = "1994"
}

@article{ATLAS:2024erm,
    author = "Aad, Georges and others",
    collaboration = "ATLAS",
    title = "{Measurement of the W-boson mass and width with the ATLAS detector using proton{\textendash}proton collisions at $\sqrt{s}=7$ TeV}",
    eprint = "2403.15085",
    archivePrefix = "arXiv",
    primaryClass = "hep-ex",
    reportNumber = "CERN-EP-2024-074",
    doi = "10.1140/epjc/s10052-024-13190-x",
    journal = "Eur. Phys. J. C",
    volume = "84",
    number = "12",
    pages = "1309",
    year = "2024"
}

@article{ATLAS:2023lhg,
    author = "Aad, Georges and others",
    collaboration = "ATLAS",
    title = "{A precise determination of the strong-coupling constant from the recoil of $Z$ bosons with the ATLAS experiment at $\sqrt{s} = 8$ TeV}",
    eprint = "2309.12986",
    archivePrefix = "arXiv",
    primaryClass = "hep-ex",
    month = "9",
    year = "2023"
}

@article{Mueller:1984vh,
    author = "Mueller, Alfred H.",
    title = "{On the Structure of Infrared Renormalons in Physical Processes at High-Energies}",
    reportNumber = "CU-TP-289",
    doi = "10.1016/0550-3213(85)90485-7",
    journal = "Nucl. Phys. B",
    volume = "250",
    pages = "327--350",
    year = "1985"
}

@article{Parisi:1978bj,
    author = "Parisi, G.",
    title = "{Singularities of the Borel Transform in Renormalizable Theories}",
    reportNumber = "LPTENS 78/8",
    doi = "10.1016/0370-2693(78)90101-6",
    journal = "Phys. Lett. B",
    volume = "76",
    pages = "65--66",
    year = "1978"
}

@article{Gross:1974jv,
    author = "Gross, David J. and Neveu, Andre",
    title = "{Dynamical Symmetry Breaking in Asymptotically Free Field Theories}",
    reportNumber = "COO-2220-19",
    doi = "10.1103/PhysRevD.10.3235",
    journal = "Phys. Rev. D",
    volume = "10",
    pages = "3235",
    year = "1974"
}

@article{Lautrup:1977hs,
    author = "Lautrup, B. E.",
    title = "{On High Order Estimates in QED}",
    reportNumber = "NBI-HE-77-6",
    doi = "10.1016/0370-2693(77)90145-9",
    journal = "Phys. Lett. B",
    volume = "69",
    pages = "109--111",
    year = "1977"
}

@article{Beneke:1994rs,
    author = "Beneke, M.",
    title = "{More on ambiguities in the pole mass}",
    eprint = "hep-ph/9408380",
    archivePrefix = "arXiv",
    reportNumber = "UM-TH-94-31",
    doi = "10.1016/0370-2693(94)01505-7",
    journal = "Phys. Lett. B",
    volume = "344",
    pages = "341--347",
    year = "1995"
}

@article{tHooft:1977xjm,
    author = "'t Hooft, Gerard",
    editor = "Zichichi, Antonino",
    title = "{Can We Make Sense Out of Quantum Chromodynamics?}",
    reportNumber = "PRINT-77-0723 (UTRECHT)",
    journal = "Subnucl. Ser.",
    volume = "15",
    pages = "943",
    year = "1979"
}

@article{Webber:1983if,
    author = "Webber, B. R.",
    title = "{A QCD Model for Jet Fragmentation Including Soft Gluon Interference}",
    reportNumber = "CERN-TH-3713",
    doi = "10.1016/0550-3213(84)90333-X",
    journal = "Nucl. Phys. B",
    volume = "238",
    pages = "492--528",
    year = "1984"
}

@article{Sjostrand:1986hx,
    author = "Sjostrand, Torbjorn and Bengtsson, Mats",
    title = "{The Lund Monte Carlo for Jet Fragmentation and e+ e- Physics. Jetset Version 6.3: An Update}",
    reportNumber = "LU-TP-86-22",
    doi = "10.1016/0010-4655(87)90054-3",
    journal = "Comput. Phys. Commun.",
    volume = "43",
    pages = "367",
    year = "1987"
}

@article{Sjostrand:1982fn,
    author = "Sjostrand, Torbjorn",
    title = "{The Lund Monte Carlo for Jet Fragmentation}",
    reportNumber = "LU TP 82-3",
    doi = "10.1016/0010-4655(82)90175-8",
    journal = "Comput. Phys. Commun.",
    volume = "27",
    pages = "243",
    year = "1982"
}

@article{Andersson:1983ia,
    author = "Andersson, Bo and Gustafson, G. and Ingelman, G. and Sjostrand, T.",
    title = "{Parton Fragmentation and String Dynamics}",
    reportNumber = "LU-TP-83-10",
    doi = "10.1016/0370-1573(83)90080-7",
    journal = "Phys. Rept.",
    volume = "97",
    pages = "31--145",
    year = "1983"
}

@article{Benitez:2024nav,
    author = "Benitez, Miguel A. and Hoang, Andre H. and Mateu, Vicent and Stewart, Iain W. and Vita, Gherardo",
    title = "{On determining $\alpha_{\rm s}(m_{\rm Z})$ from dijets in $e^+e^-$ thrust}",
    eprint = "2412.15164",
    archivePrefix = "arXiv",
    primaryClass = "hep-ph",
    reportNumber = "MIT-CTP 5746, CERN-TH-2024-142, UWThPh2024-8",
    doi = "10.1007/JHEP07(2025)249",
    journal = "JHEP",
    volume = "07",
    pages = "249",
    year = "2025"
}

@article{ATLAS:2024dxp,
    author = "Hayrapetyan, Aram and others",
    collaboration = "ATLAS, CMS",
    title = "{Combination of Measurements of the Top Quark Mass from Data Collected by the ATLAS and CMS Experiments at s=7 and 8~TeV}",
    eprint = "2402.08713",
    archivePrefix = "arXiv",
    primaryClass = "hep-ex",
    reportNumber = "CMS-TOP-22-001, ATLAS-TOPQ-2019-13, CERN-EP-2024-020",
    doi = "10.1103/PhysRevLett.132.261902",
    journal = "Phys. Rev. Lett.",
    volume = "132",
    number = "26",
    pages = "261902",
    year = "2024"
}

@article{Ellis:2016jkw,
    author = "Ellis, Joshua",
    title = "{TikZ-Feynman: Feynman diagrams with TikZ}",
    eprint = "1601.05437",
    archivePrefix = "arXiv",
    primaryClass = "hep-ph",
    doi = "10.1016/j.cpc.2016.08.019",
    journal = "Comput. Phys. Commun.",
    volume = "210",
    pages = "103--123",
    year = "2017"
}

@article{Lee:2013mka,
    author = "Lee, Roman N.",
    editor = "Wang, Jianxiong",
    title = "{LiteRed 1.4: a powerful tool for reduction of multiloop integrals}",
    eprint = "1310.1145",
    archivePrefix = "arXiv",
    primaryClass = "hep-ph",
    doi = "10.1088/1742-6596/523/1/012059",
    journal = "J. Phys. Conf. Ser.",
    volume = "523",
    pages = "012059",
    year = "2014"
}

@article{Dokshitzer:1995qm,
    author = "Dokshitzer, Yuri L. and Marchesini, G. and Webber, B. R.",
    title = "{Dispersive approach to power behaved contributions in QCD hard processes}",
    eprint = "hep-ph/9512336",
    archivePrefix = "arXiv",
    reportNumber = "CERN-TH-95-281, CAVENDISH-HEP-95-12",
    doi = "10.1016/0550-3213(96)00155-1",
    journal = "Nucl. Phys. B",
    volume = "469",
    pages = "93--142",
    year = "1996"
}

@article{Dasgupta:1999zm,
    author = "Dasgupta, Mrinal",
    title = "{Power corrections to the differential Drell-Yan cross-section}",
    eprint = "hep-ph/9911391",
    archivePrefix = "arXiv",
    reportNumber = "BICOCCA-FT-99-33",
    doi = "10.1088/1126-6708/1999/12/008",
    journal = "JHEP",
    volume = "12",
    pages = "008",
    year = "1999"
}

@article{Dasgupta:2007wa,
    author = "Dasgupta, Mrinal and Magnea, Lorenzo and Salam, Gavin P.",
    title = "{Non-perturbative QCD effects in jets at hadron colliders}",
    eprint = "0712.3014",
    archivePrefix = "arXiv",
    primaryClass = "hep-ph",
    reportNumber = "DFTT-27-2007, MAN-HEP-2007-41",
    doi = "10.1088/1126-6708/2008/02/055",
    journal = "JHEP",
    volume = "02",
    pages = "055",
    year = "2008"
}

@article{FerrarioRavasio:2021mzg,
    author = "Ferrario Ravasio, Silvia",
    title = "{Infrared renormalons in collider processes}",
    eprint = "2106.00276",
    archivePrefix = "arXiv",
    primaryClass = "hep-ph",
    doi = "10.1140/epjs/s11734-021-00254-2",
    journal = "Eur. Phys. J. ST",
    volume = "230",
    number = "12-13",
    pages = "2581--2592",
    year = "2021"
}

@article{Dokshitzer:1998pt,
    author = "Dokshitzer, Yuri L. and Lucenti, A. and Marchesini, G. and Salam, G. P.",
    title = "{On the universality of the Milan factor for 1 / Q power corrections to jet shapes}",
    eprint = "hep-ph/9802381",
    archivePrefix = "arXiv",
    reportNumber = "IFUM-601-FT",
    doi = "10.1088/1126-6708/1998/05/003",
    journal = "JHEP",
    volume = "05",
    pages = "003",
    year = "1998"
}

@article{Dokshitzer:1997iz,
    author = "Dokshitzer, Yuri L. and Lucenti, A. and Marchesini, G. and Salam, G. P.",
    title = "{Universality of 1/Q corrections to jet-shape observables rescued}",
    eprint = "hep-ph/9707532",
    archivePrefix = "arXiv",
    reportNumber = "IFUM-573-FT",
    doi = "10.1016/S0550-3213(97)00650-0",
    journal = "Nucl. Phys. B",
    volume = "511",
    pages = "396--418",
    year = "1998",
    note = "[Erratum: Nucl.Phys.B 593, 729--730 (2001)]"
}

@article{Dokshitzer:1997ew,
    author = "Dokshitzer, Yuri L. and Webber, B. R.",
    title = "{Power corrections to event shape distributions}",
    eprint = "hep-ph/9704298",
    archivePrefix = "arXiv",
    reportNumber = "CAVENDISH-HEP-97-2",
    doi = "10.1016/S0370-2693(97)00573-X",
    journal = "Phys. Lett. B",
    volume = "404",
    pages = "321--327",
    year = "1997"
}

@article{Beneke:1997sr,
    author = "Beneke, M. and Braun, Vladimir M. and Magnea, Lorenzo",
    title = "{Phenomenology of power corrections in fragmentation processes in e+ e- annihilation}",
    eprint = "hep-ph/9701309",
    archivePrefix = "arXiv",
    reportNumber = "CERN-TH-96-362, NORDITA-96-79-P",
    doi = "10.1016/S0550-3213(97)00251-4",
    journal = "Nucl. Phys. B",
    volume = "497",
    pages = "297--333",
    year = "1997"
}

@inproceedings{Korchemsky:1996iq,
    author = "Korchemsky, G. P.",
    title = "{Power corrections in Drell-Yan production beyond the leading order}",
    booktitle = "{28th International Conference on High-energy Physics}",
    eprint = "hep-ph/9610207",
    archivePrefix = "arXiv",
    reportNumber = "LPTHE-ORSAY-96-78",
    pages = "793--796",
    month = "10",
    year = "1996"
}

@article{Dasgupta:1996ki,
    author = "Dasgupta, M. and Webber, B. R.",
    title = "{Power corrections and renormalons in $e^{+} e^{-}$ fragmentation functions}",
    eprint = "hep-ph/9608394",
    archivePrefix = "arXiv",
    reportNumber = "CAVENDISH-HEP-96-9",
    doi = "10.1016/S0550-3213(96)00622-0",
    journal = "Nucl. Phys. B",
    volume = "484",
    pages = "247--264",
    year = "1997"
}

@article{Mizuta:1992ja,
    author = "Mizuta, Satoshi and Ng, Daniel and Yamaguchi, Masahiro",
    title = "{Phenomenological aspects of supersymmetric standard models without grand unification}",
    eprint = "hep-ph/9210241",
    archivePrefix = "arXiv",
    reportNumber = "TU-410, IFP-436-UNC",
    doi = "10.1016/0370-2693(93)90754-6",
    journal = "Phys. Lett. B",
    volume = "300",
    pages = "96--103",
    year = "1993"
}

@article{Kinoshita:1962ur,
    author = "Kinoshita, T.",
    title = "{Mass singularities of Feynman amplitudes}",
    doi = "10.1063/1.1724268",
    journal = "J. Math. Phys.",
    volume = "3",
    pages = "650--677",
    year = "1962"
}

@article{Akhoury:1995cj,
    author = "Akhoury, R. and Zakharov, Valentin I.",
    title = "{Power corrections in QCD: A Matter of energy resolution}",
    eprint = "hep-ph/9512433",
    archivePrefix = "arXiv",
    reportNumber = "UM-TH-95-33",
    doi = "10.1103/PhysRevLett.76.2238",
    journal = "Phys. Rev. Lett.",
    volume = "76",
    pages = "2238--2241",
    year = "1996"
}

@article{Bloch:1937pw,
    author = "Bloch, F. and Nordsieck, A.",
    title = "{Note on the Radiation Field of the electron}",
    reportNumber = "RX-1199",
    doi = "10.1103/PhysRev.52.54",
    journal = "Phys. Rev.",
    volume = "52",
    pages = "54--59",
    year = "1937"
}

@article{Lee:1964is,
    author = "Lee, T. D. and Nauenberg, M.",
    editor = "Feinberg, G.",
    title = "{Degenerate Systems and Mass Singularities}",
    doi = "10.1103/PhysRev.133.B1549",
    journal = "Phys. Rev.",
    volume = "133",
    pages = "B1549--B1562",
    year = "1964"
}

@article{Shtabovenko:2016whf,
    author = "Shtabovenko, Vladyslav",
    title = "{FeynHelpers: Connecting FeynCalc to FIRE and Package-X}",
    eprint = "1611.06793",
    archivePrefix = "arXiv",
    primaryClass = "physics.comp-ph",
    reportNumber = "TUM-EFT-75-15",
    doi = "10.1016/j.cpc.2017.04.014",
    journal = "Comput. Phys. Commun.",
    volume = "218",
    pages = "48--65",
    year = "2017"
}

@article{Beneke:1998ui,
    author = "Beneke, M.",
    title = "{Renormalons}",
    eprint = "hep-ph/9807443",
    archivePrefix = "arXiv",
    reportNumber = "CERN-TH-98-233",
    doi = "10.1016/S0370-1573(98)00130-6",
    journal = "Phys. Rept.",
    volume = "317",
    pages = "1--142",
    year = "1999"
}

@article{FerrarioRavasio:2018ubr,
    author = "Ferrario Ravasio, Silvia and Nason, Paolo and Oleari, Carlo",
    title = "{All-orders behaviour and renormalons in top-mass observables}",
    eprint = "1810.10931",
    archivePrefix = "arXiv",
    primaryClass = "hep-ph",
    doi = "10.1007/JHEP01(2019)203",
    journal = "JHEP",
    volume = "01",
    pages = "203",
    year = "2019"
}

@article{Nason:1995np,
    author = "Nason, Paolo and Seymour, Michael H.",
    title = "{Infrared renormalons and power suppressed effects in e+ e- jet events}",
    eprint = "hep-ph/9506317",
    archivePrefix = "arXiv",
    reportNumber = "CERN-TH-95-150, IFUM-507-FT",
    doi = "10.1016/0550-3213(95)00461-Z",
    journal = "Nucl. Phys. B",
    volume = "454",
    pages = "291--312",
    year = "1995"
}

@article{Dokshitzer:1995zt,
    author = "Dokshitzer, Yuri L. and Webber, B. R.",
    title = "{Calculation of power corrections to hadronic event shapes}",
    eprint = "hep-ph/9504219",
    archivePrefix = "arXiv",
    reportNumber = "CAVENDISH-HEP-95-2, LU-TP-95-8",
    doi = "10.1016/0370-2693(95)00548-Y",
    journal = "Phys. Lett. B",
    volume = "352",
    pages = "451--455",
    year = "1995"
}

@article{Akhoury:1995sp,
    author = "Akhoury, R. and Zakharov, Valentin I.",
    title = "{On the universality of the leading, 1/Q power corrections in QCD}",
    eprint = "hep-ph/9504248",
    archivePrefix = "arXiv",
    reportNumber = "SACLAY-SPH-T-95-043, UM-TH-95-12",
    doi = "10.1016/0370-2693(95)00866-J",
    journal = "Phys. Lett. B",
    volume = "357",
    pages = "646--652",
    year = "1995"
}

@article{Beneke:1995pq,
    author = "Beneke, M. and Braun, Vladimir M.",
    title = "{Power corrections and renormalons in Drell-Yan production}",
    eprint = "hep-ph/9506452",
    archivePrefix = "arXiv",
    reportNumber = "DESY-95-120, UM-TH-95-17",
    doi = "10.1016/0550-3213(95)00439-Y",
    journal = "Nucl. Phys. B",
    volume = "454",
    pages = "253--290",
    year = "1995"
}

@article{Ball:1995ni,
    author = "Ball, Patricia and Beneke, M. and Braun, Vladimir M.",
    title = "{Resummation of (beta0 alpha-s)**n corrections in QCD: Techniques and applications to the tau hadronic width and the heavy quark pole mass}",
    eprint = "hep-ph/9502300",
    archivePrefix = "arXiv",
    reportNumber = "CERN-TH-95-26, CERN-TH-95-026, UM-TH-95-3",
    doi = "10.1016/0550-3213(95)00392-6",
    journal = "Nucl. Phys. B",
    volume = "452",
    pages = "563--625",
    year = "1995"
}

@article{Beneke:1994bc,
      author         = "Beneke, M. and Braun, Vladimir M. and Zakharov, Valentin
                        I.",
      title          = "{Bloch-Nordsieck cancellations beyond logarithms in heavy
                        particle decays}",
      journal        = "Phys. Rev. Lett.",
      volume         = "73",
      year           = "1994",
      pages          = "3058-3061",
      doi            = "10.1103/PhysRevLett.73.3058",
      eprint         = "hep-ph/9405304",
      archivePrefix  = "arXiv",
      primaryClass   = "hep-ph",
      reportNumber   = "MPI-PHT-94-18, UM-TH-94-13",
      SLACcitation   = "%%CITATION = HEP-PH/9405304;%%"
}

@article{Makarov:2023ttq,
    author = "Makarov, Sergei and Melnikov, Kirill and Nason, Paolo and Ozcelik, Melih A.",
    title = "{Linear power corrections to single top production processes at the LHC}",
    eprint = "2302.02729",
    archivePrefix = "arXiv",
    primaryClass = "hep-ph",
    reportNumber = "TTP23-003, P3H-23-007",
    doi = "10.1007/JHEP05(2023)153",
    journal = "JHEP",
    volume = "05",
    pages = "153",
    year = "2023"
}

@article{Denner:2019vbn,
    author = "Denner, Ansgar and Dittmaier, Stefan",
    title = "{Electroweak Radiative Corrections for Collider Physics}",
    eprint = "1912.06823",
    archivePrefix = "arXiv",
    primaryClass = "hep-ph",
    reportNumber = "FR-PHENO-019",
    doi = "10.1016/j.physrep.2020.04.001",
    journal = "Phys. Rept.",
    volume = "864",
    pages = "1--163",
    year = "2020"
}

@article{Ross:1973fp,
    author = "Ross, D. A. and Taylor, J. C.",
    title = "{Renormalization of a unified theory of weak and electromagnetic interactions}",
    doi = "10.1016/0550-3213(73)90505-1",
    journal = "Nucl. Phys. B",
    volume = "51",
    pages = "125--144",
    year = "1973",
    note = "[Erratum: Nucl.Phys.B 58, 643--643 (1973)]"
}

@article{FerrarioRavasio:2020guj,
    author = "Ferrario Ravasio, Silvia and Limatola, Giovanni and Nason, Paolo",
    title = "{Infrared renormalons in kinematic distributions for hadron collider processes}",
    eprint = "2011.14114",
    archivePrefix = "arXiv",
    primaryClass = "hep-ph",
    reportNumber = "OUTP-20-13P, IPPP/20/60",
    doi = "10.1007/JHEP06(2021)018",
    journal = "JHEP",
    volume = "06",
    pages = "018",
    year = "2021"
}

@article{Makarov:2023uet,
    author = "Makarov, Sergei and Melnikov, Kirill and Nason, Paolo and Ozcelik, Melih A.",
    title = "{Linear power corrections to top quark pair production in hadron collisions}",
    eprint = "2308.05526",
    archivePrefix = "arXiv",
    primaryClass = "hep-ph",
    reportNumber = "TTP23-032, P3H-23-055",
    doi = "10.1007/JHEP01(2024)074",
    journal = "JHEP",
    volume = "01",
    pages = "074",
    year = "2024"
}

@article{Caola:2022vea,
    author = "Caola, Fabrizio and Ferrario Ravasio, Silvia and Limatola, Giovanni and Melnikov, Kirill and Nason, Paolo and Ozcelik, Melih Arslan",
    title = "{Linear power corrections to e$^{+}$e$^{–}$ shape variables in the three-jet region}",
    eprint = "2204.02247",
    archivePrefix = "arXiv",
    primaryClass = "hep-ph",
    reportNumber = "OUTP-22-04P, TTP22-022, P3H-22-036, MPP-2022-36",
    doi = "10.1007/JHEP12(2022)062",
    journal = "JHEP",
    volume = "12",
    pages = "062",
    year = "2022"
}

@article{Gherghetta:1999sw,
    author = "Gherghetta, Tony and Giudice, Gian F. and Wells, James D.",
    title = "{Phenomenological consequences of supersymmetry with anomaly induced masses}",
    eprint = "hep-ph/9904378",
    archivePrefix = "arXiv",
    reportNumber = "CERN-TH-99-104",
    doi = "10.1016/S0550-3213(99)00429-0",
    journal = "Nucl. Phys. B",
    volume = "559",
    pages = "27--47",
    year = "1999"
}

@article{Yamada:2009ve,
    author = "Yamada, Youichi",
    title = "{Electroweak two-loop contribution to the mass splitting within a new heavy SU(2)(L) fermion multiplet}",
    eprint = "0906.5207",
    archivePrefix = "arXiv",
    primaryClass = "hep-ph",
    reportNumber = "TU-848",
    doi = "10.1016/j.physletb.2009.11.044",
    journal = "Phys. Lett. B",
    volume = "682",
    pages = "435--440",
    year = "2010"
}

@article{Czarnecki:1996nr,
    author = "Czarnecki, Andrzej and Smirnov, Vladimir A.",
    title = "{Threshold behavior of Feynman diagrams: The Master two loop propagator}",
    eprint = "hep-ph/9608407",
    archivePrefix = "arXiv",
    reportNumber = "TTP-96-32",
    doi = "10.1016/S0370-2693(96)01698-X",
    journal = "Phys. Lett. B",
    volume = "394",
    pages = "211--217",
    year = "1997"
}

@article{Glashow:1961tr,
    author = "Glashow, S. L.",
    title = "{Partial Symmetries of Weak Interactions}",
    doi = "10.1016/0029-5582(61)90469-2",
    journal = "Nucl. Phys.",
    volume = "22",
    pages = "579--588",
    year = "1961"
}

@article{Salam:1968rm,
    author = "Salam, Abdus",
    title = "{Weak and Electromagnetic Interactions}",
    doi = "10.1142/9789812795915_0034",
    journal = "Conf. Proc. C",
    volume = "680519",
    pages = "367--377",
    year = "1968"
}

@article{Weinberg:1967tq,
    author = "Weinberg, Steven",
    title = "{A Model of Leptons}",
    doi = "10.1103/PhysRevLett.19.1264",
    journal = "Phys. Rev. Lett.",
    volume = "19",
    pages = "1264--1266",
    year = "1967"
}

@article{Bigi:1997fj,
    author = "Bigi, Ikaros I. Y. and Shifman, Mikhail A. and Uraltsev, N.",
    title = "{Aspects of heavy quark theory}",
    eprint = "hep-ph/9703290",
    archivePrefix = "arXiv",
    reportNumber = "TPI-MINN-97-02-T, UMN-TH-1528-97, UND-HEP-97-BIG01",
    doi = "10.1146/annurev.nucl.47.1.591",
    journal = "Ann. Rev. Nucl. Part. Sci.",
    volume = "47",
    pages = "591--661",
    year = "1997"
}

@article{Chetyrkin:1997dh,
    author = "Chetyrkin, K. G.",
    title = "{Quark mass anomalous dimension to O (alpha-s**4)}",
    eprint = "hep-ph/9703278",
    archivePrefix = "arXiv",
    reportNumber = "MPI-PHT-97-019",
    doi = "10.1016/S0370-2693(97)00535-2",
    journal = "Phys. Lett. B",
    volume = "404",
    pages = "161--165",
    year = "1997"
}

@article{Chetyrkin:1981qh,
    author = "Chetyrkin, K. G. and Tkachov, F. V.",
    title = "{Integration by parts: The algorithm to calculate $\beta$-functions in 4 loops}",
    doi = "10.1016/0550-3213(81)90199-1",
    journal = "Nucl. Phys. B",
    volume = "192",
    pages = "159--204",
    year = "1981"
}

@article{Becher:2008cf,
    author = "Becher, Thomas and Schwartz, Matthew D.",
    title = "{A precise determination of $\alpha_s$ from LEP thrust data using effective field theory}",
    eprint = "0803.0342",
    archivePrefix = "arXiv",
    primaryClass = "hep-ph",
    reportNumber = "FERMILAB-PUB-08-048-T",
    doi = "10.1088/1126-6708/2008/07/034",
    journal = "JHEP",
    volume = "07",
    pages = "034",
    year = "2008"
}

@article{Bell:2023dqs,
    author = "Bell, Guido and Lee, Christopher and Makris, Yiannis and Talbert, Jim and Yan, Bin",
    title = "{Effects of renormalon scheme and perturbative scale choices on determinations of the strong coupling from e+e- event shapes}",
    eprint = "2311.03990",
    archivePrefix = "arXiv",
    primaryClass = "hep-ph",
    doi = "10.1103/PhysRevD.109.094008",
    journal = "Phys. Rev. D",
    volume = "109",
    number = "9",
    pages = "094008",
    year = "2024"
}

@article{Nason:2025qbx,
    author = "Nason, Paolo and Zanderighi, Giulia",
    title = "{Fits of {\ensuremath{\alpha}}$_{s}$ from event-shapes in the three-jet region: extension to all energies}",
    eprint = "2501.18173",
    archivePrefix = "arXiv",
    primaryClass = "hep-ph",
    reportNumber = "MPP-2025-9",
    doi = "10.1007/JHEP06(2025)200",
    journal = "JHEP",
    volume = "06",
    pages = "200",
    year = "2025"
}

@article{Nason:2023asn,
    author = "Nason, Paolo and Zanderighi, Giulia",
    title = "{Fits of {\ensuremath{\alpha}}$_{s}$ using power corrections in the three-jet region}",
    eprint = "2301.03607",
    archivePrefix = "arXiv",
    primaryClass = "hep-ph",
    doi = "10.1007/JHEP06(2023)058",
    journal = "JHEP",
    volume = "06",
    pages = "058",
    year = "2023"
}

@article{Camarda:2022qdg,
    author = "Camarda, Stefano and Ferrera, Giancarlo and Schott, Matthias",
    title = "{Determination of the strong-coupling constant from the Z-boson transverse-momentum distribution}",
    eprint = "2203.05394",
    archivePrefix = "arXiv",
    primaryClass = "hep-ph",
    doi = "10.1140/epjc/s10052-023-12373-2",
    journal = "Eur. Phys. J. C",
    volume = "84",
    number = "1",
    pages = "39",
    year = "2024"
}

@article{Hoang:1998hm,
    author = "Hoang, Andre H. and Ligeti, Zoltan and Manohar, Aneesh V.",
    title = "{B decays in the upsilon expansion}",
    eprint = "hep-ph/9811239",
    archivePrefix = "arXiv",
    reportNumber = "UCSD-PTH-98-32, CERN-TH-98-334, FERMILAB-PUB-98-351-T",
    doi = "10.1103/PhysRevD.59.074017",
    journal = "Phys. Rev. D",
    volume = "59",
    pages = "074017",
    year = "1999"
}

@article{Hoang:2018zrp,
    author = {Hoang, Andr{\'e} H. and Pl{\"a}tzer, Simon and Samitz, Daniel},
    title = "{On the Cutoff Dependence of the Quark Mass Parameter in Angular Ordered Parton Showers}",
    eprint = "1807.06617",
    archivePrefix = "arXiv",
    primaryClass = "hep-ph",
    reportNumber = "UWTHPH-2018-20, MCnet-18-17",
    doi = "10.1007/JHEP10(2018)200",
    journal = "JHEP",
    volume = "10",
    pages = "200",
    year = "2018"
}

@article{Shifman:1978bx,
    author = "Shifman, Mikhail A. and Vainshtein, A. I. and Zakharov, Valentin I.",
    title = "{QCD and Resonance Physics. Theoretical Foundations}",
    reportNumber = "ITEP-73-1978, ITEP-80-1978",
    doi = "10.1016/0550-3213(79)90022-1",
    journal = "Nucl. Phys. B",
    volume = "147",
    pages = "385--447",
    year = "1979"
}

@article{Wilson:1969zs,
    author = "Wilson, Kenneth G.",
    title = "{Nonlagrangian models of current algebra}",
    doi = "10.1103/PhysRev.179.1499",
    journal = "Phys. Rev.",
    volume = "179",
    pages = "1499--1512",
    year = "1969"
}

\end{document}